\documentclass[11pt,a4paper]{article}
\pdfoutput=1 

\usepackage{jheppub} 
\usepackage[T1]{fontenc}
\usepackage{bbm}
\usepackage{amssymb,amsmath,amsfonts}
\usepackage{mathrsfs,mathtools}
\usepackage[all]{hypcap}
\usepackage{stackrel}
\usepackage{float}

\renewcommand{\thefigure}{\Roman{figure}}

\newcommand{\be}{\begin{equation}}
\newcommand{\ee}{\end{equation}}
\newcommand{\bse}{\begin{subequations}}
\newcommand{\ese}{\end{subequations}}
\newcommand{\ba}{\begin{eqnarray}}
\newcommand{\ea}{\end{eqnarray}}
\renewcommand{\(}{\left(}
\renewcommand{\)}{\right)}

\newcommand{\lk}{\left[}
\newcommand{\rk}{\right]}

\newcommand{\s}{\sigma}

\DeclareMathOperator{\extdm}{d}
\newcommand{\extd}{\extdm \!}

\begin{document}

\title{Evolution of holographic entanglement entropy in an anisotropic system}

\author[]{Christian Ecker, Daniel Grumiller and Stefan A.~Stricker}

\affiliation[]{Institut f\"{u}r Theoretische Physik, Technische Universit\"{a}t Wien,\\
Wiedner Hauptstr.~8-10, A-1040 Vienna, Austria}

\emailAdd{ecker@hep.itp.tuwien.ac.at}
\emailAdd{grumil@hep.itp.tuwien.ac.at}
\emailAdd{stricker@hep.itp.tuwien.ac.at}

\abstract{
We determine holographically 2-point correlators of gauge invariant operators with large conformal weights and entanglement entropy of strips for a time-dependent anisotropic 5-dimensional asymptotically anti-de~Sitter spacetime. At the early stage of evolution where geodesics and extremal surfaces can extend beyond the apparent horizon all observables vary substantially from their thermal value, but thermalize rapidly. At late times we recover quasi-normal ringing of correlators and holographic entanglement entropy around their thermal values, as expected on general grounds. We check the behaviour of holographic entanglement entropy and correlators as function of the separation length of the strip and find agreement with the exact expressions derived in the small and large temperature limits.
}

\rightline{\href{http://www.itp.tuwien.ac.at/Fundamental_Interactions\#2015}{TUW--15--12}}
\maketitle

\listoffigures

\section{Introduction}
 Far from equilibrium dynamics and thermalization  of strongly coupled systems has  attracted a lot of attention in the past decade. The reason for this is twofold. 
 On the one hand experiments at RHIC and the LHC revealed that the quark gluon plasma created in heavy ion collisions behaves as a strongly coupled liquid that thermalizes extremely fast \cite{Teaney:2000cw,Huovinen:2001cy,Hirano:2002ds,Romatschke:2007mq}. 
 On the other hand,  in condensed matter experiments it is now possible to drive an isolated system  to a far from equilibrium state by a quantum quench, i.e.~a control  parameter of the system is varied rapidly \cite{Polkovnikov:2010yn}. 
  
 One powerful tool to study strongly coupled systems out of equilibrium is the gauge gravity duality where classical supergravity in $d+1$ dimensions is dual to a $d$-dimensional strongly coupled large-$N$ field theory that loosely speaking lives on the boundary of the gravity theory \cite{Maldacena:1997re, Witten:1998qj, Gubser:1998bc}. The thermalization process of field theories with a conformal field theory fixed point in the ultraviolet is then  mapped to black hole (or black brane) formation in asymptotically Anti-de Sitter (AdS) space \cite{Witten:1998zw}.

Particularly useful theoretical observables to monitor the thermalization process are the vacuum expectation value of the stress-energy tensor, correlation functions of gauge invariant operators and entanglement entropy (EE). The former two are easy to define on the field theory side and, given some standard assumptions that we recall below, easy to calculate on the gravity side. We recall now relevant aspects of EE, see \cite{Eisert:2008ur} for more details and references.

On the field theory side EE is defined in the following way. Dividing a system into two subsystems $A$ and $B$ the EE $S_A$  of the subsystem $A$ is defined as the von Neumann entropy of the  reduced density matrix obtained by tracing out the degrees of freedom of the subsystem $B$.

On the gravity side, the holographic entanglement entropy (HEE) is defined as the area of a minimal surface extending from some predefined surface $A$ on the boundary into the bulk \cite{Ryu:2006bv,Nishioka:2009un}. 
For time-dependent backgrounds the minimal surface has to be replaced by an extremal surface \cite{Hubeny:2007xt}. It is noteworthy --- both for conceptual and for technical reasons --- that the relevant surfaces used to determine HEE can extend beyond the apparent horizon. Conceptually, this implies that HEE provides a `classical' way (on the gravity side) to extract (quantum) information from the region beyond the horizon. Technically, this requires the numerical determination of (part of) the spacetime region beyond the apparent horizon, which can be a challenge since that region ends in a singularity.

EE is notoriously hard to compute in quantum field theories; so far this is only possible in highly symmetric theories such as (relativistic) conformal field theories \cite{Calabrese:2004eu,Holzhey:1994we,Vidal:2002rm} or Galilean conformal field theories \cite{Bagchi:2014iea}.
In a seminal work, Calabrese and Cardy \cite{Calabrese:2005in} were able to compute the time evolution of the EE after a quench in a  two dimensional conformal field theory and in the Ising spin chain  model in a transverse magnetic field. In both cases they find that for an entangling interval of length $l$, the EE increases linearly with time until $t\sim l/2$ after which it saturates. The linear scaling with time and the crossover at $t\sim l/2$ can be understood in terms of entangled  quasiparticles pairs emitted from the initial state and is therefore  expected to hold for a wider class of systems. 

The holographic duality offers a playground where one can study various different systems and search for universal and novel properties of HEE in dynamical situations. 

The simplest example where one can study the analog of quenches in the holographic setup are spacetimes where thin shells collapse to form a black hole, first utilised in  \cite{AbajoArrastia:2010yt}. 
The behaviour  of the EE in these setups has been studied extensively \cite{AbajoArrastia:2010yt, Balasubramanian:2011ur, Albash:2010mv, Baron:2012fv,Galante:2012pv,Keranen:2011xs, Liu:2013qca} and indeed shows universal behaviour consistent with the findings of Cardy and Calabrese. 
Namely the  initial short  early time epoch  is followed by a long period where the EE grows linearly with  time and is independent of the entangling boundary region. 
 This was first worked out in the Vaidya spacetime  \cite{AbajoArrastia:2010yt, Liu:2013iza, Liu:2013qca} where the shell is composed out of null dust and later generalized to matter with  arbitrary equation of states \cite{Keranen:2014zoa, Keranen:2015fqa}. The linear  scaling is even present in geometries with Lifshitz scaling and hyper scaling violation \cite{Alishahiha:2014cwa, Fonda:2014ula}. 
 In addition, in the above works the HEE is a monotonically increasing function that saturates to its equilibrium value from below. 
 Note that these quenches are thermal quenches because the end state is a thermal state.

Another way of studying  thermal quenches is by turning on  a radially collapsing scalar field that forms a black hole. This situation is more complicated because in order to obtain the geometry from which one can then obtain the HEE,  Einsteins equations have to be solved numerically \cite{Bizon:2011gg, Abajo-Arrastia:2014fma, Buchel:2014gta}. 
In \cite{Abajo-Arrastia:2014fma,daSilva:2014zva} this was done for a radial collapsing massless scalar field in  global AdS where the scalar field can have many bounces between the boundary and the center of AdS before a black hole forms resulting  in a periodic behaviour of the HEE.

In \cite{Buchel:2014gta}  a massive scalar field dual to a massive fermionic operator was turned on, treating the quench as a perturbation on the static spacetime. In this setup the HEE is also not monotonic and in some cases approaches the equilibrium value from above. This reveals a qualitative difference of EE to the thermal entropy which, on general grounds, must be monotonically growing in a closed system.

An additional motivation to study HEE comes from the question how to measure entropy production in (holographic models for) heavy ion collisions \cite{Muller:2011ra}. Within the gauge/gravity duality the entropy of the (stationary) black hole corresponds to the entropy of the field theory. However, in time-dependent backgrounds entropy as defined from the area of the apparent horizon is ambiguous because it depends on the choice of time slicing. By contrast the definition of the HEE is unique and therefore may serve as an alternative measure for entropy production.  

So far nearly all studies of HEE have relied on the simplifying 
assumption of a spherically symmetric spacetime (see \cite{Narayan:2012ks} 
for a notable exception).
In the paper at hand we drop this assumption and investigate the effect of anisotropies on the evolution of the HEE.

The paper is organized at follows. 
In section \ref{se:2} we introduce the anisotropic spacetime we will use, recall general aspects of 2-point correlators and simplify the determination of HEE to a geodesic problem in an auxiliary spacetime.
In section \ref{se:numerics} we discuss the numerical strategy that we used, relegating details to the appendices.
In section \ref{se:3} we present results for the background geometry, holographic stress tensor, 2-point correlators and HEE.
In section \ref{se:4} we analyze the late time behaviour of correlators and HEE and relate it to the lowest lying quasinormal mode.
In section \ref{se:5} we conclude with a brief summary of our results and some possible next steps. 
In appendix \ref{se:spectral} we comment on the spectral method and other numerical routines used to solve the 5-dimensional vacuum Einstein equations.
In appendix \ref{se:relax} we explain the relaxation code used to determine geodesics and extremal surfaces.

Before starting we mention some of our conventions. We use mostly $+$ signature and set the AdS radius and final black brane mass to unity.

\section{Theoretical setup}\label{se:2}

\subsection{Anisotropic asymptotically AdS$_5$ spacetimes}\label{se:2.1}

In this section we review the most important details of the model first introduced in \cite{Chesler:2008hg} and studied further in 
\cite{Chesler:2009cy, Heller:2012km, Heller:2013oxa, Chesler:2013lia}. 

The 5-dimensional bulk metric that introduces anisotropy between the longitudinal and transverse directions with an $\mathcal{O}(2)$ rotational invariance in the transverse plane can be written conveniently in Eddington--Finkelstein coordinates,

\be\label{metric}
\extd s^2=-A(r,v)\extd v^2+2\extd r\extd v +\Sigma ^2(r,v)\Big( e^{-2B(r,v)} \extd x_\parallel^2 +e^{B(r,v)} \extd\vec{x}_\perp^2 \Big)
\ee
where the functions $A,~B$ and $\Sigma$ only depend on the holographic coordinate $r$ and (advanced) time $v$. 
In this coordinate system the vacuum Einstein equations read 
\bse
\label{einstein}
\ba
0 &=& \Sigma (\dot{\Sigma})'+2\Sigma' \dot{\Sigma}-2\Sigma^2 \label{eq:E1}\\
0 &=&2\Sigma(\dot{B})'+3(\Sigma'\dot{B}+B'\dot{\Sigma}) \label{eq:E2} \\
0&=&A''+ 3B'\dot{B}-12 \Sigma' \dot{\Sigma}/\Sigma^2+4 \label{eq:E3}\\
0&=& 2\ddot{\Sigma}-A' \dot{\Sigma}+\dot{B}^2\Sigma \label{eq:E4}\\
0&=& 2\Sigma'' + (B')^2\Sigma \label{eq:E5}
\ea
\ese
where prime denotes radial derivative and dot time derivative, viz.
\be\label{dot}
h'\equiv \partial_r h\qquad \dot{h}\equiv\partial_v h+\frac{1}{2}A \partial_r h
\ee
for any function $h(r,v)$.

The Einstein equations \eqref{einstein} have to be solved for special initial conditions and appropriate  boundary conditions. 
There are, at least, two ways to create a far from equilibrium state. 
On the one hand one can turn on a time dependent anisotropy function at the boundary $B(r=\infty,v)=B_0(t)$ as in the original works of \cite{Chesler:2008hg, Chesler:2009cy} and let the system evolve. In this case the boundary metric is curved and the conformal anomaly is present \cite{Henningson:1998gx}. 
On the other hand one can specify the initial state in the absence of external sources by specifying the metric in the bulk on the initial time slice \cite{Heller:2012km} with a flat boundary geometry. 
For simplicity, in the following  we will study the setup where the boundary metric is flat and time independent. 

Requiring that the spacetime is asymptotically AdS$_5$,
 Einstein's equations in the near boundary expansion $(r\rightarrow\infty)$ are solved by\footnote{%
We fix a residual gauge freedom that preserves the form \eqref{metric}, $r\to r + f(v)$, by demanding that the first subleading term in the function $A$ falls off like ${\cal O}(r^{-2})$. Note that in several numerical simulations in the literature the same freedom is fixed by placing the apparent horizon at a specific value of the radial coordinate $r$. 
 } 
\bse
\ba
A&=&r^2+\frac{a_4}{r^2}-\frac{2b_4(t)^2}{7r^6}+\mathcal{O}(r^{-7})\\
B&=&\frac{b_4(t)}{r^4}+\frac{\partial_t b_4(t)}{r^5}+\mathcal{O}(r^{-6})\\
\Sigma&=&r-\frac{b_4(t)^2}{7r^7}+\mathcal{O}(r^{-8})\,. 
\ea
\ese
The coefficients in the asymptotic expansion determine the expectation value of the stress energy tensor in the dual field theory \cite{deHaro:2000xn}
\be
\langle T^{\mu\nu}\rangle =\frac{N_c^2}{2\pi^2}\,\mathrm{diag}\lk \mathcal{E},~P_\parallel (t),~P_\perp (t),~P_\perp (t) \rk
\ee
where 
\be
\mathcal{E} = -\frac{3}{4}a_4\qquad P_\parallel(t) = -\frac{1}{4}a_4 -2 b_4(t)\qquad P_\perp(t) = -\frac{1}{4}a_4 +b_4(t)\,.
\ee
In order to determine the fourth order coefficients one needs to solve Einstein's equations numerically for some initial conditions. When  doing the actual calculation we work with the inverse radial coordinate $z=1/r$ so that the boundary is located at $z=0$. 

For our initial data  we follow \cite{Heller:2013oxa, Chesler:2013lia} and choose for the anisotropy function on the initial time slice 
\be
B(r,v_0)=\frac{\beta}{r^4}\exp \lk-\Big(\frac{1}{r}-\frac{1}{r_0}\Big)^2/\omega^2\rk 
\label{eq:initial}
\ee
with $\beta=6.6$, $r_0=4$ and $w=1$.
In addition  the initial conditions have to be supplemented  with a value for the coefficient $a_4$ which sets the energy density of the initial state, for which we take  $a_4=-1$,  corresponding to an equilibrium temperature $T=1/\pi$, as we now recall.

At late times we expect isotropization, $B=0$. In that case we recover the usual static AdS black brane solution as follows. Solving \eqref{eq:E5} and using residual gauge transformations yields $\Sigma=r$. This implies $\Sigma^\prime=1$ and $\dot\Sigma=\tfrac12\, A$. Solving \eqref{eq:E1} then yields $A=r^2\,(1-1/r^4)$, where we fixed the integration constant such that $a_4=-1$. [The other equations are either trivial, \eqref{eq:E2} and \eqref{eq:E4}, or redundant, \eqref{eq:E3}.] The result for $A$ is the usual Killing norm for the static AdS black brane. Surface gravity is given by $\kappa = \tfrac12\,A^\prime\big|_{r=1}=2$ so that the Hawking temperature is $T=\kappa/(2\pi)=1/\pi$.

In the generic anisotropic case, $B\neq 0$, we solve the Einstein equations \eqref{einstein} numerically for the initial conditions \eqref{eq:initial}.
In this background we then study the evolution of 2-point correlation functions for operators of large conformal weights and the HEE. This in turn requires us to determine the background sufficiently far beyond the apparent horizon. In section \ref{se:3.1} below we discuss the numerical implementation.

\subsection{2-point correlators}\label{se:2.2}

The equal time 2-point function for an operator of large conformal weight $\Delta$ can be computed via a path integral as \cite{Balasubramanian:1999zv, Festuccia:2005pi}
\be
\langle \mathcal{O}(t, \vec{x})\mathcal{O}(t, \vec{x}')\rangle=\int  \mathcal{D P}\, e^{i \Delta \mathcal{L(\mathcal{P})} }\approx \!\!\!\sum_{\textrm{\tiny geodesics}}\!\!\! e^{-\Delta L_g}\approx e^{-\Delta L}
\ee
where the  integral is a sum over all possible paths with endpoints at $(t, \vec{x}')$ and $(t, \vec{x})$ and $\mathcal{L(P)}$ is the proper length of the path.  The first approximation neglects perturbative corrections and is the so called geodesic approximation, which holds in the limit when the conformal weight of the operator is large. The conformal weight effectively plays the role of $1/\hbar$ in usual perturbative expansions of path integrals. Then it can be shown that the sum over all paths reduces to a sum over all geodesics where $L_g$ denotes the length of the corresponding geodesic. To leading order only the geodesic with the smallest value of $L_g$ contributes, whose length we denote by $L$, which explains the second approximation\footnote{
For a comparison of the  2-point correlation function obtained by  using the ``extrapolate'' dictionary and the geodesic approximation in AdS$_3$ Vaidya spacetime see \cite{Keranen:2014lna}.}. It neglects instanton corrections.

However, the length of the geodesic has a divergence originating from the asymptotically AdS boundary and therefore needs to be  renormalized. We choose to subtract the length of a geodesic in the static  black brane background, which we denote by $L_{\textrm{\tiny therm}}$. In terms of the renormalized length $\delta L=L-L_{\textrm{\tiny therm}}$ the 2-point function becomes
\be
\langle \mathcal{O}(t, \vec{x})\mathcal{O}(t, \vec{x}')\rangle\sim e^{-\Delta \delta L}\,.
\ee 
This means that we can obtain  the time evolution of 2-point functions by looking at spacelike  geodesics that are anchored at the boundary at fixed separation $l$ and calculating their length at different times.
Due to the anisotropy in the system  we only solve for the subset of correlation functions that are either separated in the longitudinal direction or in the transverse directions.

To this end we let all the coordinates depend on one parameter $\sigma$, which lies in the interval $\s\in [-\s_m,~ \s_m]$.
To obtain the lengths of the geodesics we have to solve the geodesic equation for the two subspaces given by the line elements 
\ba
\extd s^2_\perp&=& -A \extd v^2-\frac{2}{z^2}  \extd z \extd v +\Sigma^2 e^{B}\extd x_\perp^2 \\
\extd s^2_\parallel&=&  -A \extd v^2-\frac{2}{z^2}  \extd z \extd v +\Sigma^2 e^{-2B} \extd x_\parallel^2\,.
\ea

For the separation in the transverse direction the geodesics end at $(v(\pm \s_m)=t, \allowbreak \,x_{\perp_1}(\pm \s_m)=\pm x_0/2, \, x_{\perp_2}(\pm \s_m)=0, \,x_\parallel (\pm \s_m)=0)$, where $t$ is the boundary time.
Similarly, for the longitudinal separation we take  $(v(\pm \s_m)=t,~x_{\parallel}(\pm \s_m)=\pm x_0/2,~\vec{x}_{\perp}(\pm \s_m)=0)$. 
With this choice of boundary conditions the lengths of the geodesics in the background (\ref{metric}) are given by
\ba
L_\perp&=&\int_{-\s_m}^{\s_m} \extd\s \sqrt{-A (v')^2-\frac{2}{z^2} z'v'+\Sigma^2 e^{B}(x'_\perp) ^2}\\
L_\parallel&=&\int_{-\s_m}^{\s_m} \extd\s \sqrt{-A (v')^2-\frac{2}{z^2} z'v'+\Sigma^2 e^{-2B}(x'_\parallel) ^2}
\ea
where prime denotes the derivative with respect to $\sigma$.

It is important to point out that we can only study geodesics after some advanced time $v>v_{\textrm{\tiny min}}$ with  boundary separations  below a maximal separation $l<l_{\textrm{\tiny max}}$. 
This comes from the fact that by solving Einstein's equations numerically we have to choose a finite computational domain. 
Also, by  specifying  the initial state in the entire bulk on the initial time slice  the advanced time interval at our disposal is $v\in [v_0, \infty]$. 
As the geodesics reach into the bulk they bend back in advanced time leaving the computational domain for advanced times $v<v_{\textrm{\tiny min}}$ as well as extending too far into the bulk for separations $l>l_{\textrm{\tiny max}}$. 

In section \ref{se:3.2} below we discuss how to solve the geodesic equations numerically.

\subsection{Holographic entanglement entropy}\label{se:2.3}

In time dependent systems the covariant HEE \cite{Hubeny:2007xt} for some boundary region $A$  is obtained  by extremizing the 3-surface functional  
\be\label{area}
{\cal A}=\int \extd^3\sigma\sqrt{\det\Big(\frac{\partial X^{\mu}}{\partial\sigma^a}\frac{\partial X^{\nu}}{\partial\sigma^b}g_{\mu\nu}\Big)}\,
\ee
that ends on the boundary surface $A$.
In the dual field theory the EE is then conjectured to be given by \cite{Ryu:2006bv,Ryu:2006ef,Hubeny:2007xt}
\be
S_{\textrm{\tiny EE}}=\frac{\cal A}{4 G_N}\,.
\ee

Usually the boundary regions of interest are either a sphere or a strip that has finite extent in one direction and infinite extent in the other two directions. 
In spacetimes with spherical symmetry  in the three spatial dimensions the problem of finding the extremal area functional \eqref{area} effectively reduces to finding geodesics. 
In our case where spherical symmetry is broken this is not the case anymore. For example, finding the extremal area for a  spherical  boundary region would require to solve nonlinear coupled partial differential equations. 
However, in the  case of a strip with finite extent either in the transverse or longitudinal direction it is  possible to reduce the problem to finding  geodesics in a suitable auxiliary spacetime, as we now demonstrate.

We introduce two scalar fields $\phi_i(x^\alpha)$ and write the line element as
\be
\extd s^2=g_{\mu\nu}\extd x^\mu \extd x^\nu =h_{\alpha \beta} \extd x^\alpha \extd x^\beta +\phi_1^2\extd x_2^2+\phi_2^2 \extd x_3^2
\ee
where $h_{\alpha\beta}$ is a 3-dimensional metric with coordinates $(v, r, x_1)$ where $x_1$ represents the coordinate we choose to have finite spatial extent, i.e.~either $x_\parallel$ or one of $x_\perp$. The remaining (non-compact) coordinates are then denoted by $x_2$, $x_3$, which we choose to be two of our three world-volume coordinates; the third one is denoted by $\sigma$. Parametrizing the 3-dimensional coordinates as $x^{\alpha}=(v(\sigma), r(\sigma), x_1(\sigma))$, the area functional \eqref{area} can be written as 
\be
{\cal A} = \int \extd x_3 \int \extd x_2\int \extd\sigma \sqrt{\phi_1^2 \phi_2^2 h_{\alpha\beta}\frac{\partial x^{\alpha}}{\partial\sigma}\frac{\partial x^{\beta}}{\partial\sigma} }\,.
\label{eq:cg}
\ee
Performing the integration over the Killing coordinates $x_2$ and $x_3$ yields a (possibly infinite) constant volume factor through which we are going to divide. Thus, instead of calculating HEE we calculate a HEE density per Killing volume. The problem of extremizing the 3-surface corresponding to a boundary region $A$ of strip-topology is then reduced to a 1-dimensional problem.

In fact, from the expression \eqref{eq:cg} on can see that the problem of finding the extremal 3-surfaces reduces to finding geodesics of the conformal metric
\be
\extd\tilde{s}=\tilde{h}_{\alpha\beta}\extd x^\alpha \extd x^\beta=\phi_1^2\phi_2^2 h_{\alpha\beta}\extd x^\alpha \extd x^\beta\;.
\ee
The 3-dimensional conformal metrics for separation in the transverse and longitudinal directions for which we have to solve the geodesic equation in our case are given by
\begin{subequations}
 \label{eq:angelinajolie}
\ba
\extd\tilde{s}^2_\perp&=&\Sigma^4 e^{-B}   \big( -A \extd v^2+2 \extd r \extd v +\Sigma^2 e^{B}\extd x_\perp^2 \big) \\
\extd\tilde{s}^2_\parallel&=&\Sigma^4 e^{2B}   \big( -A \extd v^2+2 \extd r \extd v +\Sigma^2 e^{-2B} \extd x_\parallel ^2 \big)\,.
\ea
\end{subequations}

\section{Numerical implementation}\label{se:numerics}

\subsection{Einstein equations}\label{se:3.1}

We solve the Einstein equations \eqref{einstein} using pseudo spectral methods as described in detail in \cite{Chesler:2013lia}, with the  only difference that we do not fix the location of the apparent horizon, see appendix \ref{se:spectral} for some details. 
Not fixing the apparent horizon facilitates the study of geodesics and extremal surfaces that reach behind the apparent horizon, which is of relevance for 2-point functions and HEE. For that reason we want a large computational domain in the holographic coordinate $z=1/r$. 
In all the computations we took $z\in [0,~1.6]$ with the final position of the horizon located at $z=1$.
For the time evolution it is sufficient to use a fourth order Runge Kutta method with time steps $\delta t=10^{-3}$. 
All the computations were done with the open source software GNU Octave \cite{octave:2014}.

\subsection{Geodesics}\label{se:3.2}

To compute 2-point functions we need to find curves of extremal length in a curved spacetime whose endpoints reside on fixed positions on the boundary of that spacetime.
These curves are solutions to the geodesic equation subject to boundary conditions at the endpoints.
For numerical reasons it turns out to be convenient to use a non-affine parameter $\sigma$ , where $\tau=\tau(\sigma)$ is the usual affine parametrization, $\frac{\extd X^\mu}{\extd\tau}\frac{\extd X^\nu}{\extd\tau}g_{\mu\nu}=1$.
In terms of $\sigma$ the geodesic equation reads
\be
\ddot X^\mu + \Gamma^\mu{}_{\alpha\beta} \dot X^\alpha \dot X^\beta = -J \dot X^\mu
\label{eq:geo}
\ee
where $\dot X^\mu=\frac{\extd X^\mu}{\extd\sigma}$ and $J=\frac{\extd^2\tau}{\extd\sigma^2}/\frac{\extd\tau}{\extd\sigma}$ denotes the Jacobian which originates from the change in parametrization.
This form of the geodesic equation gives us the freedom to choose parametrizations resulting in better convergence behaviour of the relaxation algorithm than the affine parametrization does. In physical terms the right hand side in \eqref{eq:geo} introduces a fictitious viscous force that enhances numerical convergence.

In our case the geodesic equation is given by a set of three coupled nonlinear ODEs of second order for the geodesic coordinates $V,Z,X$. 
This set of equations can be reduced to a set of six first order equations in terms of the geodesic coordinates and their first derivatives:
\bse\label{geodesic2}
\ba
p_V &=& \dot V\\
p_Z &=& \dot Z\\
p_X &=& \dot X\\
\dot{p}_V + \Gamma^V{}_{VV} \, p_V^2 + \Gamma^V{}_{XX} \, p_X^2 &=&- J p_V\\
\dot{p}_Z + \Gamma^Z{}_{VV} \, p_V^2 + 2 \Gamma^Z{}_{VZ} \, p_V p_Z + \Gamma^Z{}_{ZZ}\, p_Z^2 + \Gamma^Z{}_{XX}\, p_X^2 &=&- J p_Z\\
\dot{p}_X + 2 \Gamma^X{}_{VX} \,p_V p_X + 2 \Gamma^X{}_{ZX}\, p_Z p_X &=&- J p_X
\ea
\ese

This set of equations is a two-point boundary value problem, which is usually either solved with shooting methods or relaxation methods \cite{Press:2007:NRE:1403886}. We do not shoot but relax.  The geodesics obtained in appendix \ref{se:relax} serve as the initial guess for the relaxation algorithm. Since we are interested in one-parameter families of geodesics (evaluated at different constant time slices) we can take the solution for the $n^{\rm th}$ family member as initial guess for the $(n+1)^{\rm st}$ family member. More details of our implementation of the relaxation method are described in appendix \ref{se:relax}.

\subsection{Extremal surfaces}\label{se:3.3}

The same method as above works also for HEE. Namely, for our problem at hand the evaluation of extremal surfaces is reduced to the evaluation of geodesics in some auxiliary spacetime, as we have shown in section \ref{se:2.3}.

\section{Results}\label{se:3}

In this section we display and discuss our main results. In all figures where a time-axis is plotted we measure the boundary time $t$ or the bulk advanced time $v$ in units of the temperature of the final black brane, $T=1/\pi$. The separation length $l$ of 2-point functions and HEE and the corresponding boundary coordinates are given in units of $T$ as well. 
To make the approach to thermal equilibrium most transparent we use normalized  quantities for the geodesic length  $L_\mathrm{ren}$ and HEE $S_\mathrm{ren}$ defined by 
\bse\label{ren}
\ba
L_\mathrm{ren}=\frac{L-L_\mathrm{th}}{L_\mathrm{th}}\\
S_\mathrm{ren}=\frac{S-S_\mathrm{th}}{S_\mathrm{th}}
\ea
\ese
where $L$ ($S$) is the unrenormalized length (HEE) and $L_\mathrm{th}$ ($S_\mathrm{th}$) is the corresponding thermal value.

\subsection{Background geometry and holographic stress tensor}\label{se:4.1}

\begin{figure}
\begin{center}
\hspace*{-0.8truecm}\includegraphics[scale=.45]{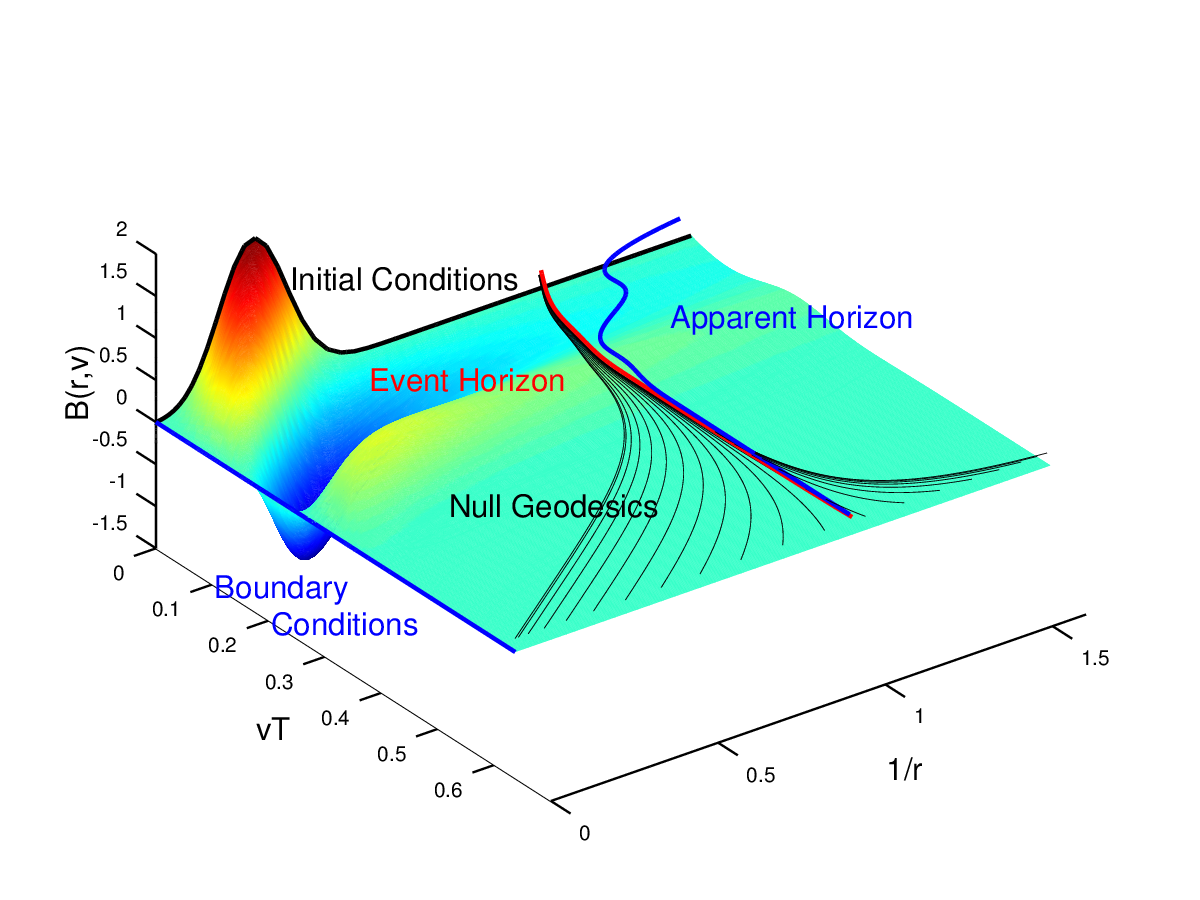}$\;\;\;\;\;$\includegraphics[scale=.35]{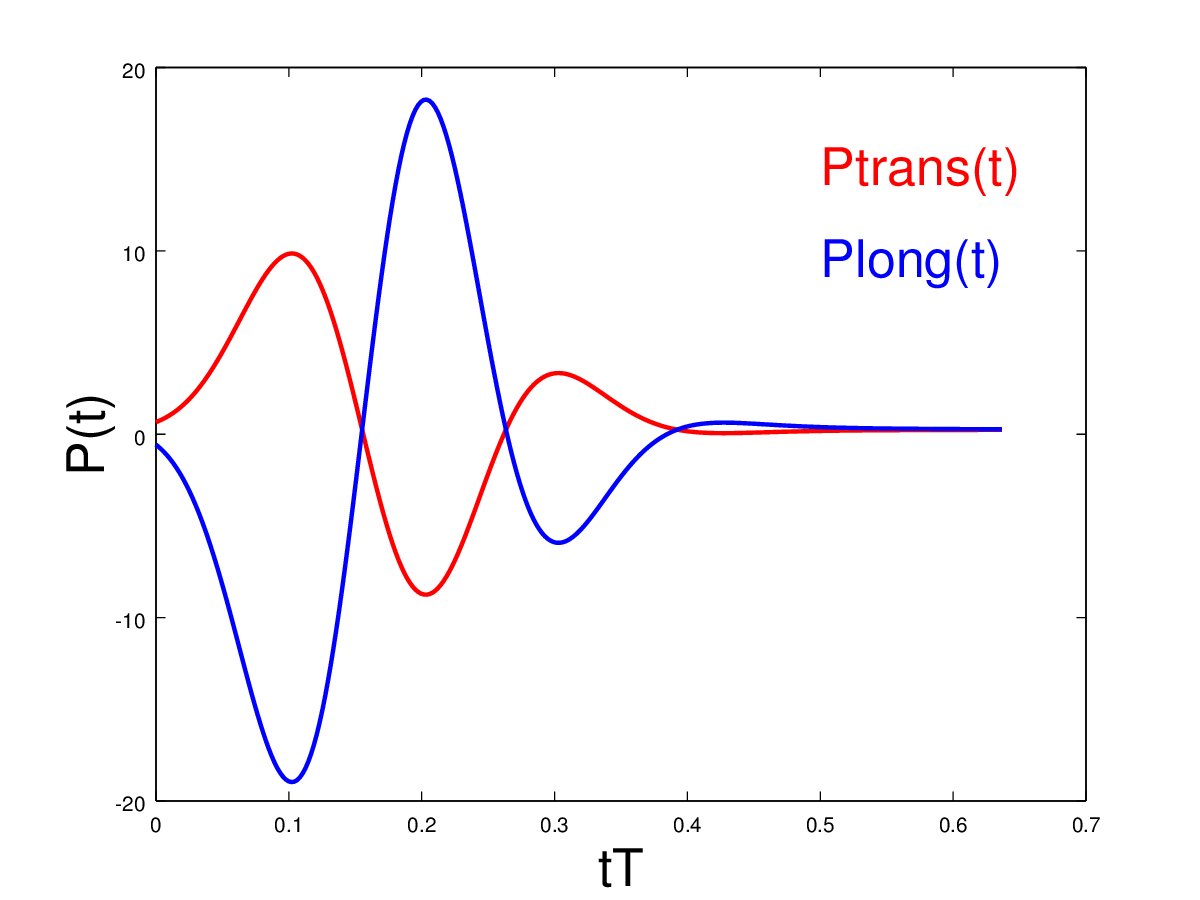}
\caption[Anisotropy function and pressures.]{\label{EMT} 
Left: anisotropy function $B(r,v)$. Right: transverse and longitudinal pressure. 
 }
\end{center}
\end{figure}

Figure \ref{EMT} displays the most salient features of the background geometry. The left figure plots the anisotropy function $B(r,v)$ and displays the regions outside and inside the apparent horizon, as well as the event horizon. The black lines depict a null congruence of geodesics close to the event horizon to exhibit their ingoing/outgoing nature. The right figure plots transversal and longitudinal pressures as function of boundary time $t$. Note the quick thermalization of the pressure components.

\subsection{2-point correlators}\label{se:4.2}

\begin{figure}
\begin{center}
\hspace*{-0.8truecm}\includegraphics[scale=.4]{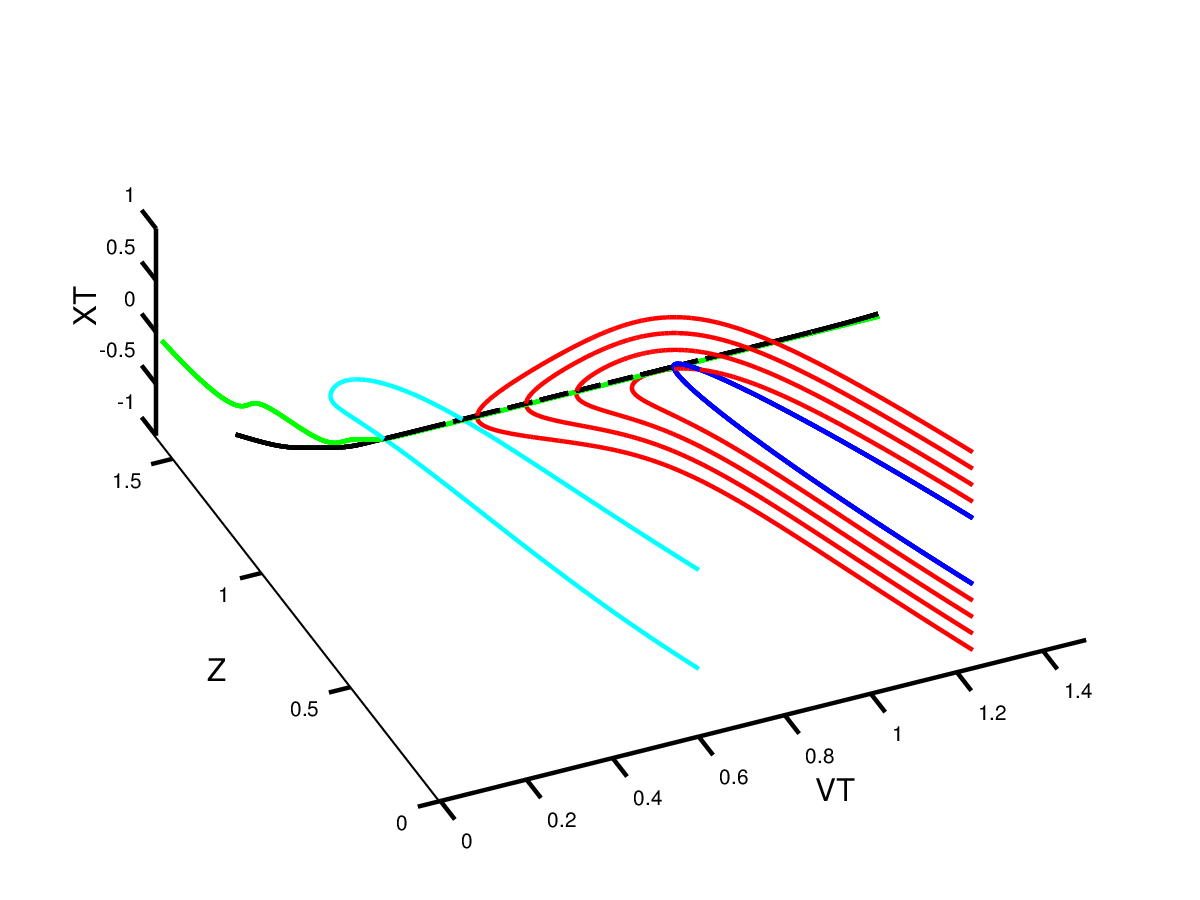}$\;\;$\includegraphics[scale=.40]{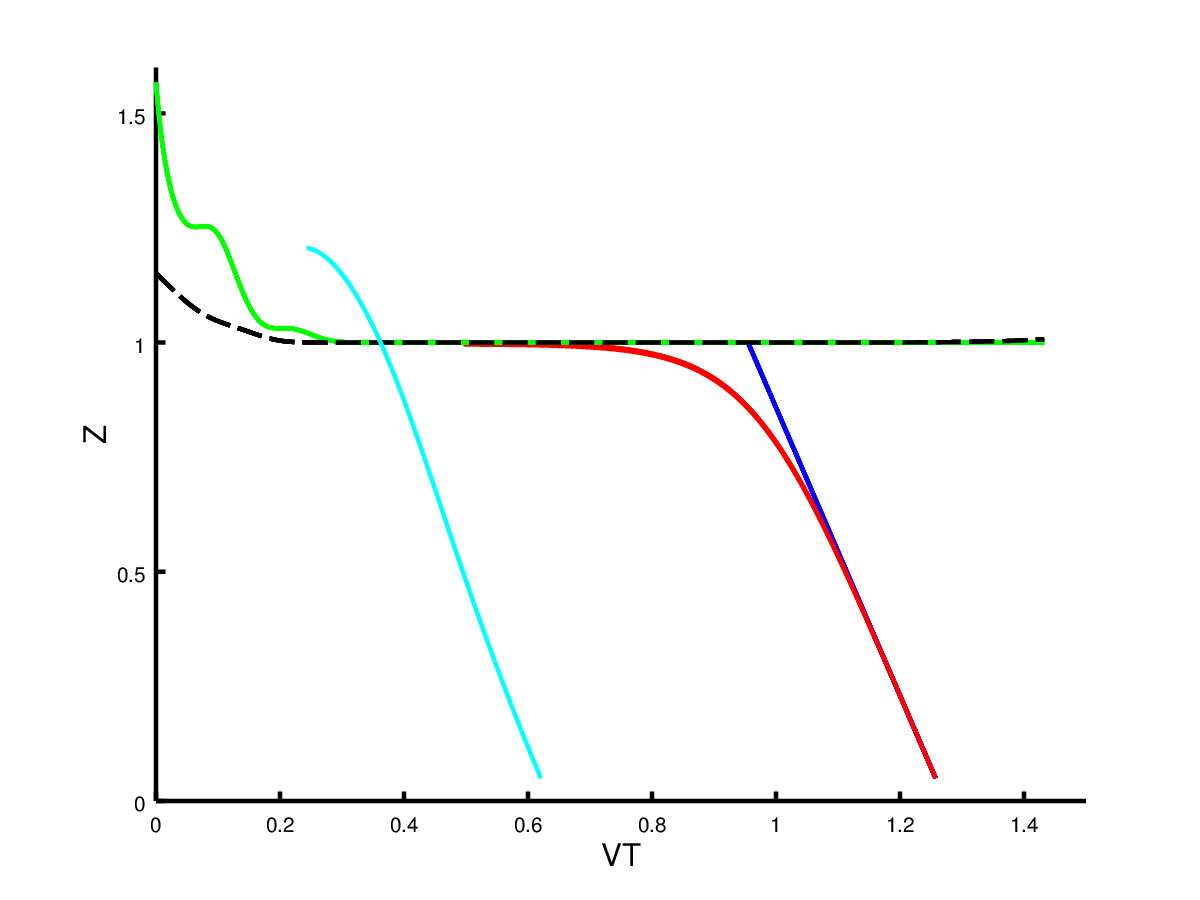}
\caption[Geodesics and their relative position to event and apparent horizon.]{\label{2PFHorizon} 
The green (black) line indicates the $z$-position of the apparent (event) horizon; 
the dark blue curve is the Poincar\'e patch AdS geodesic we use to initialize the simulation; red curves are geodesics with different boundary separation probing the thermal regime (none of them crosses the apparent horizon); the cyan curve in the left part of each plot is a geodesic which probes the non-thermal regime and reaches beyond event and apparent horizon. Isometric view (left) and view in $x$-direction (right).
 }
\end{center}
\end{figure}

As we have noted before geodesics can extend beyond the apparent horizon. This is made explicit in Fig.~\ref{2PFHorizon} where the blue curve serves as our initial guess for the relaxation code. The red geodesics at late times approach the apparent horizon without crossing it. At sufficiently early times (and sufficiently large separation) the geodesics cross the apparent horizon, an example of which is depicted by the cyan curve.

\begin{figure}
\begin{center}
\includegraphics[scale=.6]{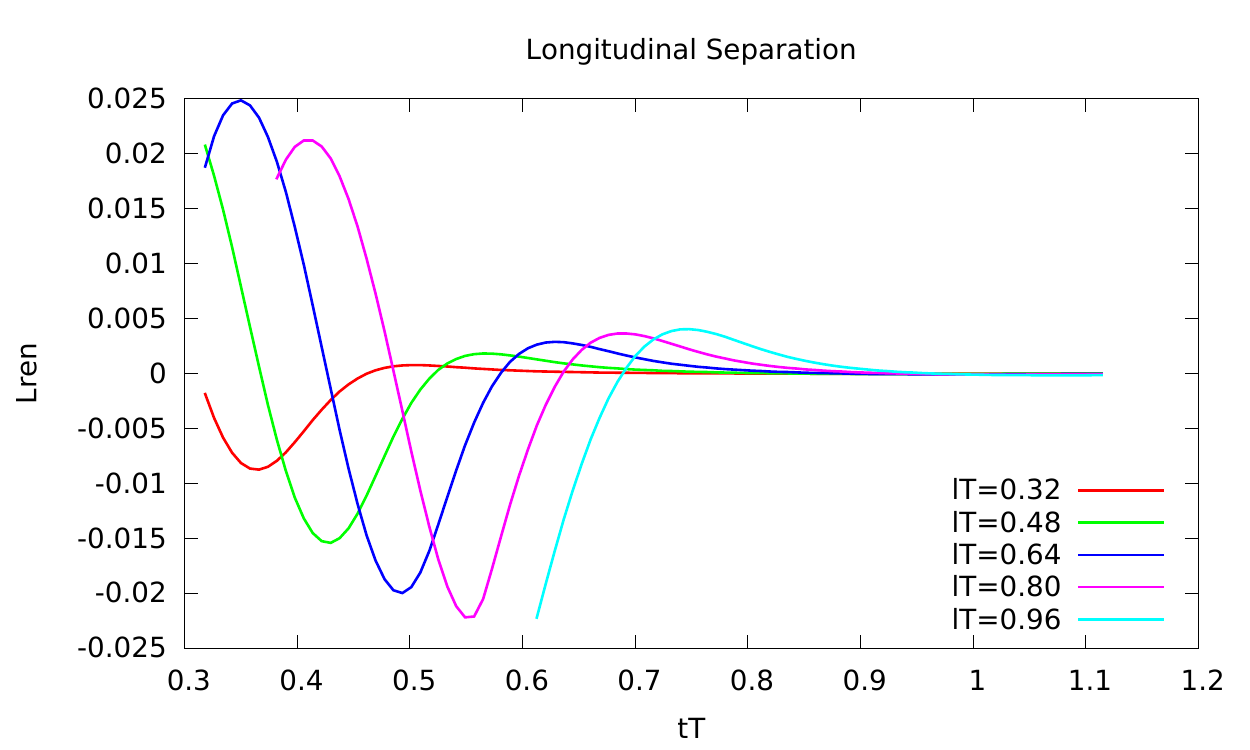}$\;\;\;\;\;$\includegraphics[scale=.60]{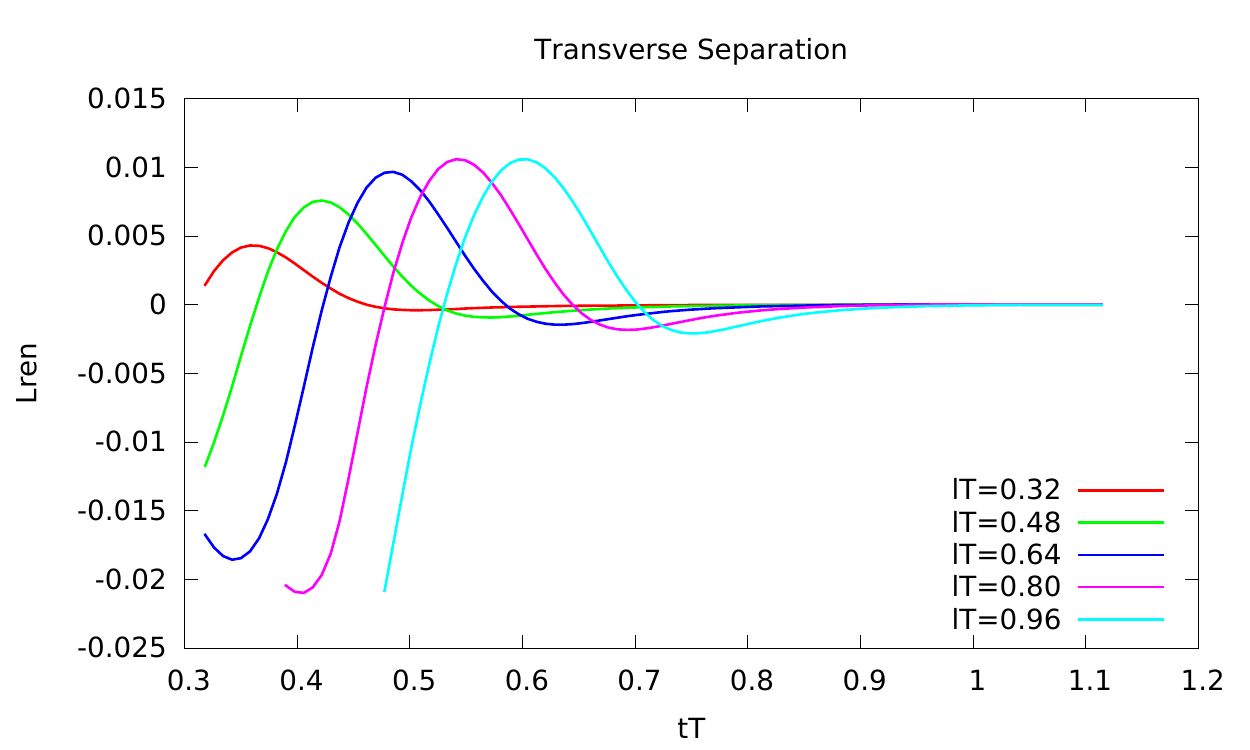}
\caption[Renormalized length of geodesics.]{\label{2PntFun} 
Renormalized length of geodesics for different separations in longitudinal and transverse directions.
 }
\end{center}
\end{figure}

The evolution of the renormalized lengths in the transverse and longitudinal directions  for different separations are  depicted in Fig.~\ref{2PntFun}. Depending on the separation the 2-point functions start at $t=t_{\rm min}$ which is the time when the geodesics extend beyond the computational domain.

\begin{figure}
\begin{center}
\includegraphics[scale=.6]{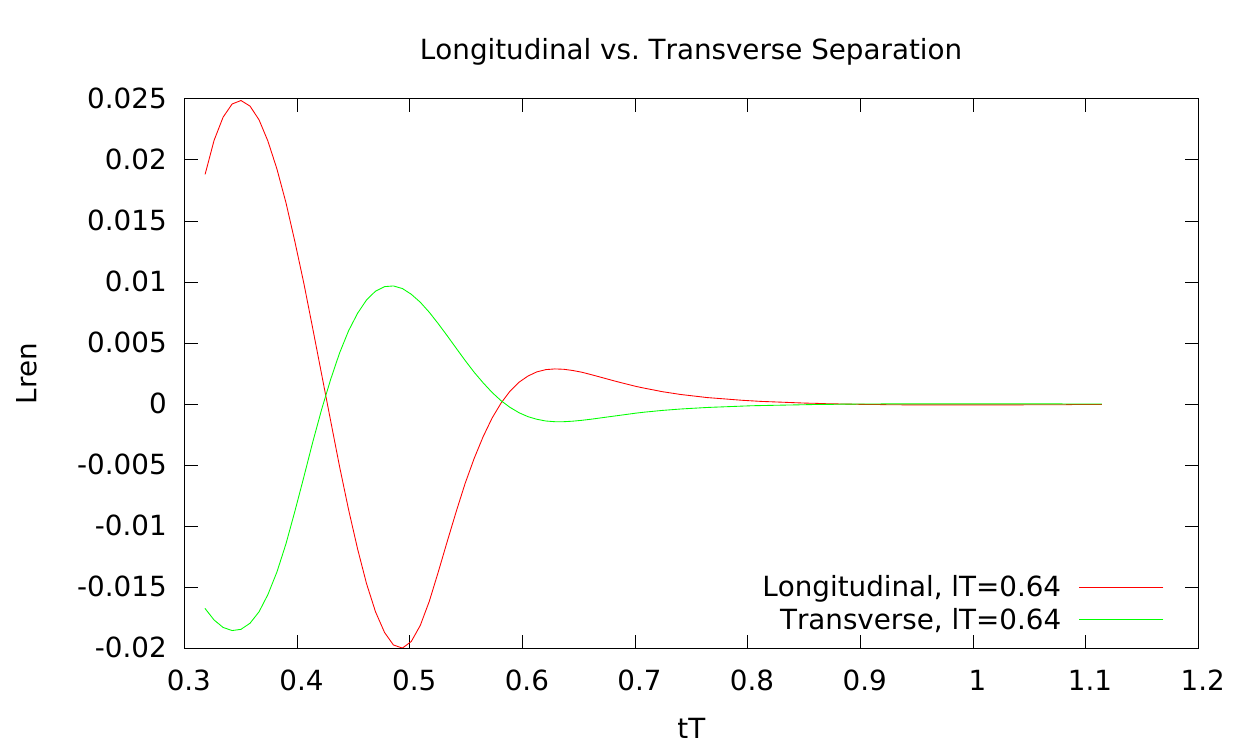}
\caption[Comparison of longitudinal and transverse geodesic lengths.]{\label{Comp2PntFun} 
Comparison of longitudinal and transverse geodesic lengths for the same boundary separation.
 }
\end{center}
\end{figure}

The first observation is that the transverse and longitudinal directions oscillate out of phase as shown in Fig.~\ref{Comp2PntFun}. The same feature is seen in the transverse and longitudinal pressure. 
By comparing the thermalization times of the one point functions, i.e.~the expectation value of the stress energy tensor with the 2-point functions we see that the 2-point functions thermalize later as expected. 
Also, the thermalization time increases if the boundary separation is increased.

\subsection{Holographic entanglement entropy}\label{se:4.3} 

The extremal surface equations --- which we mapped to geodesic equations in an auxiliary spacetime --- are solved again by a relaxation method. We observe the same qualitative features as for geodesics in Fig.~\ref{2PFHorizon} above: at early times extremal surfaces can extend beyond the apparent horizon, while at sufficiently late times they approach it from the outside without crossing. However, there are also notable differences to geodesics, which we discuss now.

As can be seen from Fig.~\ref{confgeo} in appendix \ref{app:B.2}  conformal geodesics reach much further  into the bulk  compared to the pure AdS case. 
Therefore the boundary separations we can study for  the HEE are smaller compared to the 2-point functions. 
This is also the reason why for the same boundary separation the HEE reaches equilibrium later as the 2-point functions. 
For the same boundary separation and at the same boundary time conformal geodesics reach deeper into the bulk and further back in time and therefore are more sensitive to out of equilibrium effects which are most pronounced at early times.
In addition the shape of the curves differ from the 2-point functions with the oscillations less pronounced.  We exhibit these features now in some plots.

\begin{figure}
\begin{center}
\includegraphics[scale=.6]{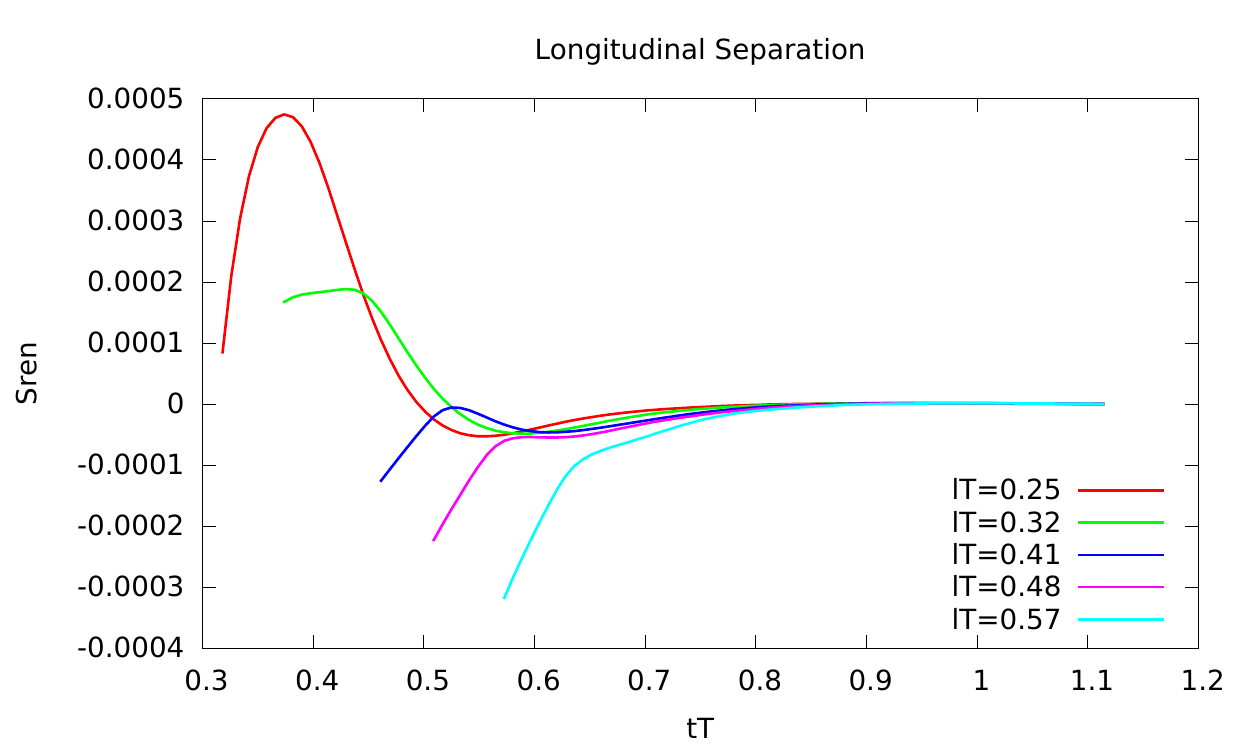}$\;\;\;\;\;$\includegraphics[scale=.6]{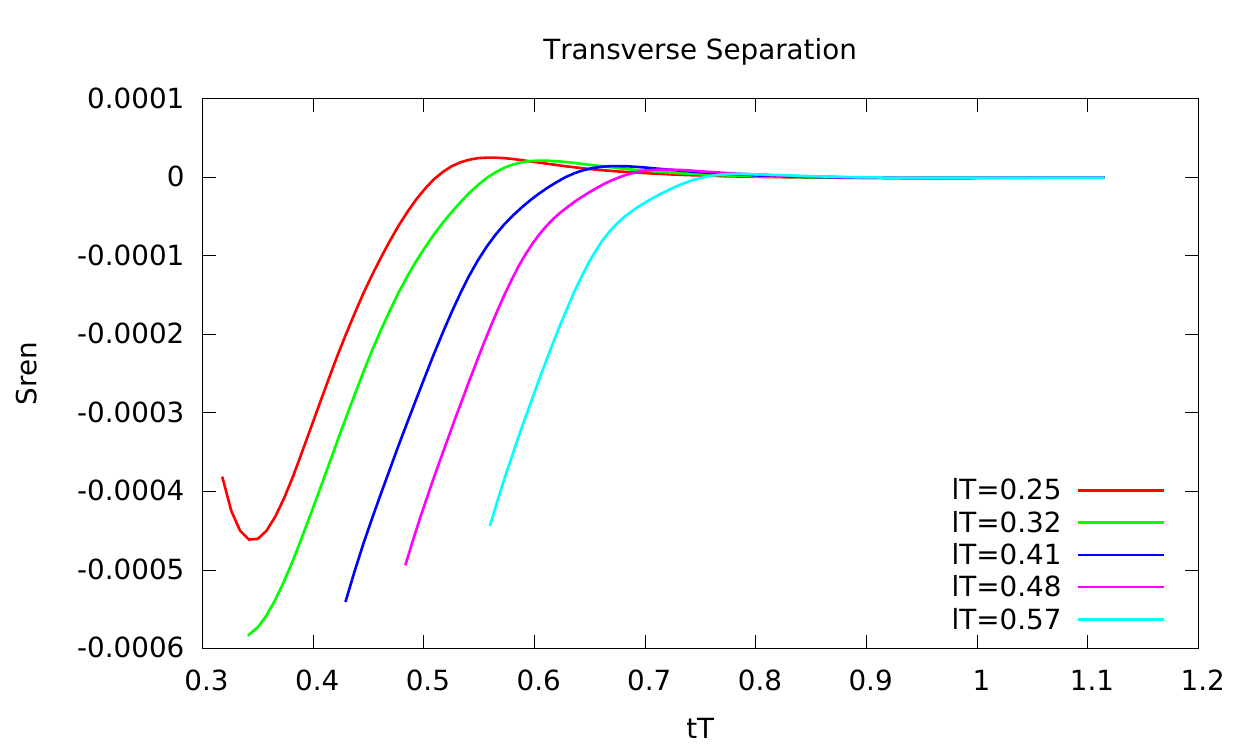}
\caption{\label{2PntFun2} 
Longitudinal and transverse HEE for different separations. 
 }
\end{center}
\end{figure}

Figure \ref{2PntFun2} plots HEE for different separations in logitudinal and transverse directions. Comparison with Fig.~\ref{2PntFun} shows that the oscillations are less pronounced for HEE.

\begin{figure}
\begin{center}
\includegraphics[scale=.6]{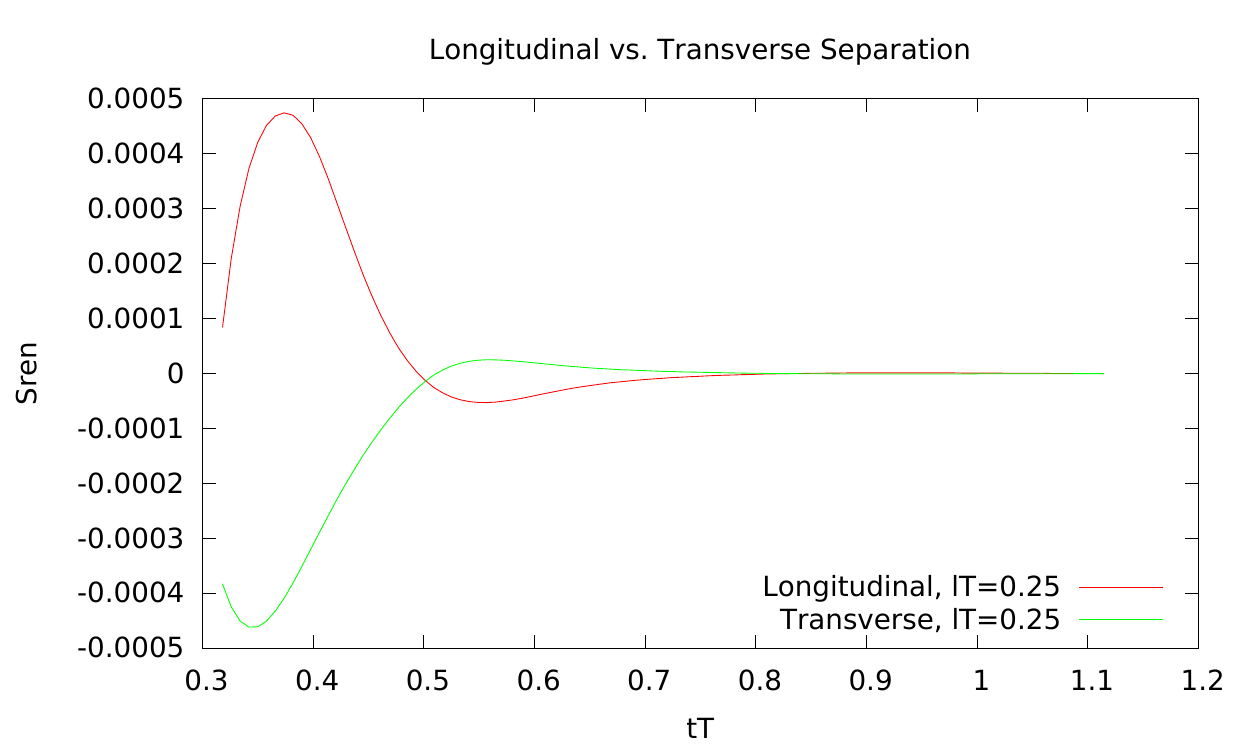}
\caption{\label{Comp2EE} 
Longitudinal and transverse HEE for same separation. 
 }
\end{center}
\end{figure}

Figure \ref{Comp2EE} plots HEE for a fixed separation in longitudinal and transverse directions. Again the behaviour of the curves is out of phase, in the sense that maxima of one curve correspond to minima of the other. Comparison with Fig.~\ref{Comp2PntFun} shows again that the oscillations are less pronounced for HEE.

\section{Late time behaviour and quasinormal modes}\label{se:4}

After the early far from equilibrium phase the geometry   relaxes to  the static Schwarzschild black brane solution. As noted in \cite{Heller:2013oxa,Chesler:2013lia} the anisotropy of the system is exponentially damped and at sufficiently late times 
one enters the  linearized regime.
In this regime the approach to equilibrium is accurately described by the lowest lying quasinormal mode (QNM) which characterizes the response of  the system  to infinitesimal metric perturbation.  
 In the case at hand the relevant channel for the gravitational fluctuations is the spin two symmetry channel which coincides with the fluctuations of a massless scalar field in the static black brane geometry.
  The asymptotic response of the pressure anisotropy then takes the form 
 
\be
b_4(t)\sim \mathrm{Re}\lk c_1 e^{-i\,\omega_1 t}\rk\,
\ee
with the lowest QNM  given by \cite{Starinets:2002br, Kovtun:2005ev} 
 \be
 \frac{\omega_1}{\pi T}=\pm 3.119452-2.746676 \,i\,.
 \ee
 On the field theory side QNMs appear as poles in the retarded Green function \cite{Birmingham:2001pj,Son:2002sd,Kovtun:2005ev,Berti:2009kk}. It is therefore expected that also the late time behaviour of the correlation functions  obtained in the previous section is described by the lowest QNM. 
We now show that this is indeed the case. 

\begin{figure}
\begin{center}
\includegraphics[scale=.6]{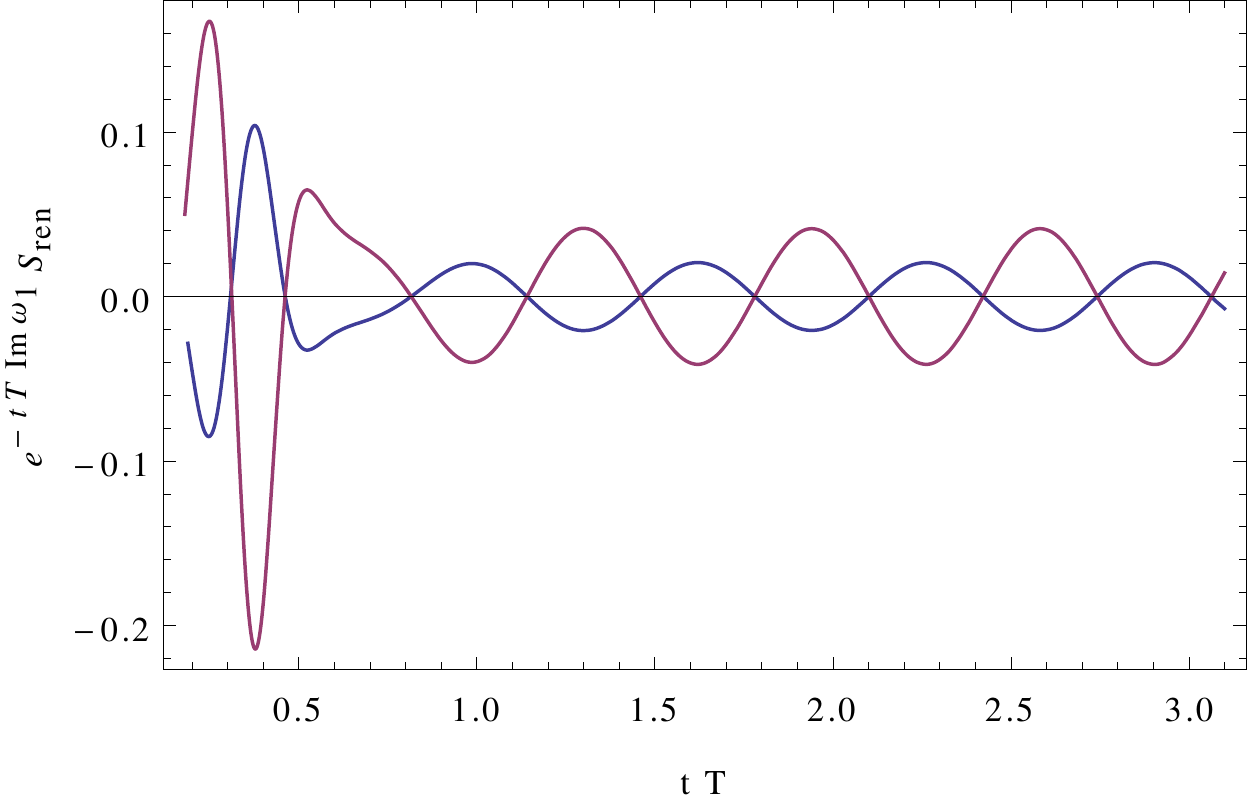}$\;\;\;\;\;$\includegraphics[scale=.6]{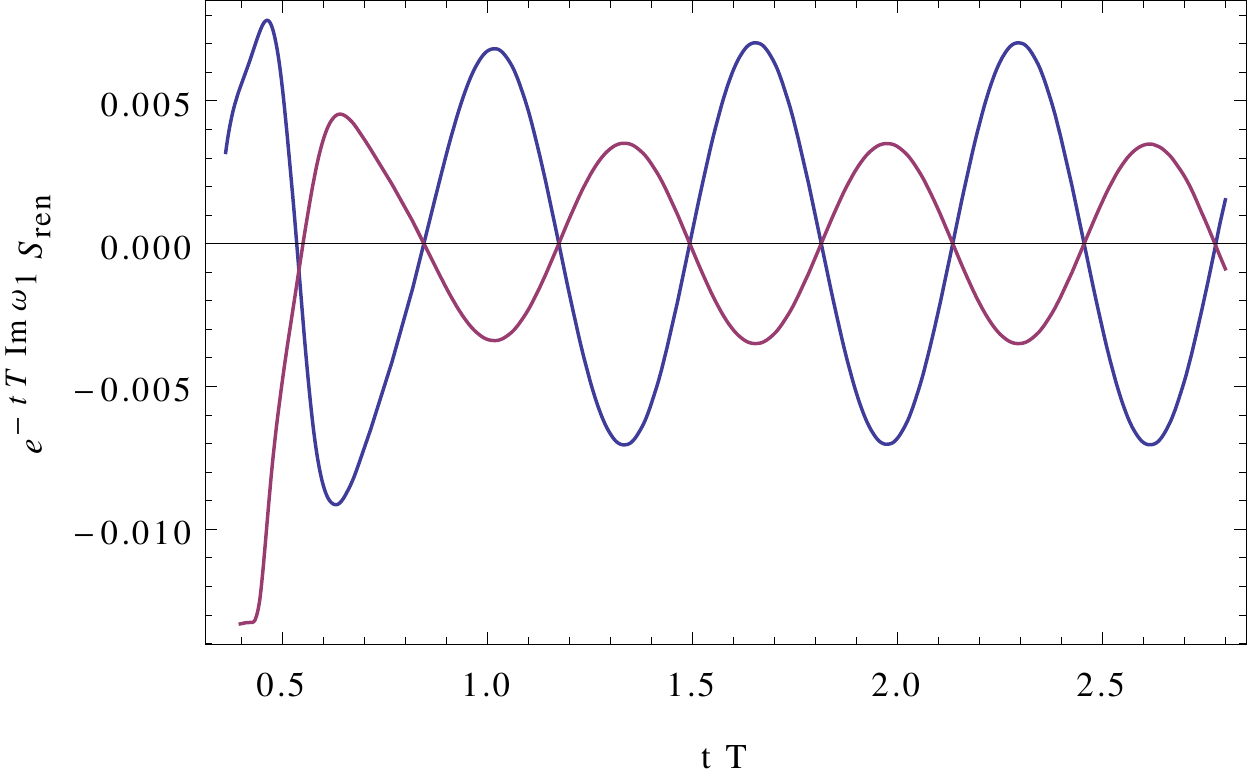}
\caption[Correlators and HEE multiplied by lowest QNM.]{\label{QNM} 
Left: Renormalized geodesic length for longitudinal (red) and transverse (blue) separation  for $lT=0.32$  multiplied by the imaginary part of the lowest QNM. Right: Renormalized HEE for the same parameters as on the left.
 }
\end{center}
\end{figure}

In figure \ref{QNM} (left) we plot the renormalized geodesic length multiplied with the imaginary part of the   lowest QNM
$e^{-\mathrm{Im}\lk\omega_1 t \rk}L_{\textrm{ren}}$ for transverse and longitudinal separations.  
One clearly sees that after a short period of time the evolution of the correlator is accurately described by the ringdown of the black brane with constant amplitude and frequency. 
The connection between the late time behaviour of correlation functions  and QNMs was previously also  observed in \cite{Balasubramanian:2012tu,Ishii:2015gia,David:2015xqa}.

It turns out that  HEE also follows this pattern. 
In \cite{Bhattacharya:2013bna} a connection between QNMs and the behaviour of the HEE was found. 
From linearised Einstein equations one can derive a differential equation for the first order correction $\Delta S_A$ of the HEE describing its change when a given ground state is excited. By imposing infalling boundary conditions  at the horizon one obtains a QNM dispersion relation putting a constraint on HEE. 
With our numerical solution we can demonstrate that the late time behaviour of HEE indeed follows the QNM ringdown even without imposing infalling boundary conditions. 
In Fig.~\ref{QNM} (right) we show the HEE multiplied with $e^{-\mathrm{Im}\lk\omega_1 t \rk} S_{\textrm{ren}}$  for the infinite strip with finite separation in longitudinal and transverse direction. As for the correlation function, at late times, the HEE shows quasinormal ringing with constant amplitude and frequency. 
These oscillations show that  HEE must not approach its thermal value from below but rather shows oscillatory behaviour around its thermal value.

\begin{figure}
\begin{center}
\includegraphics[scale=.6]{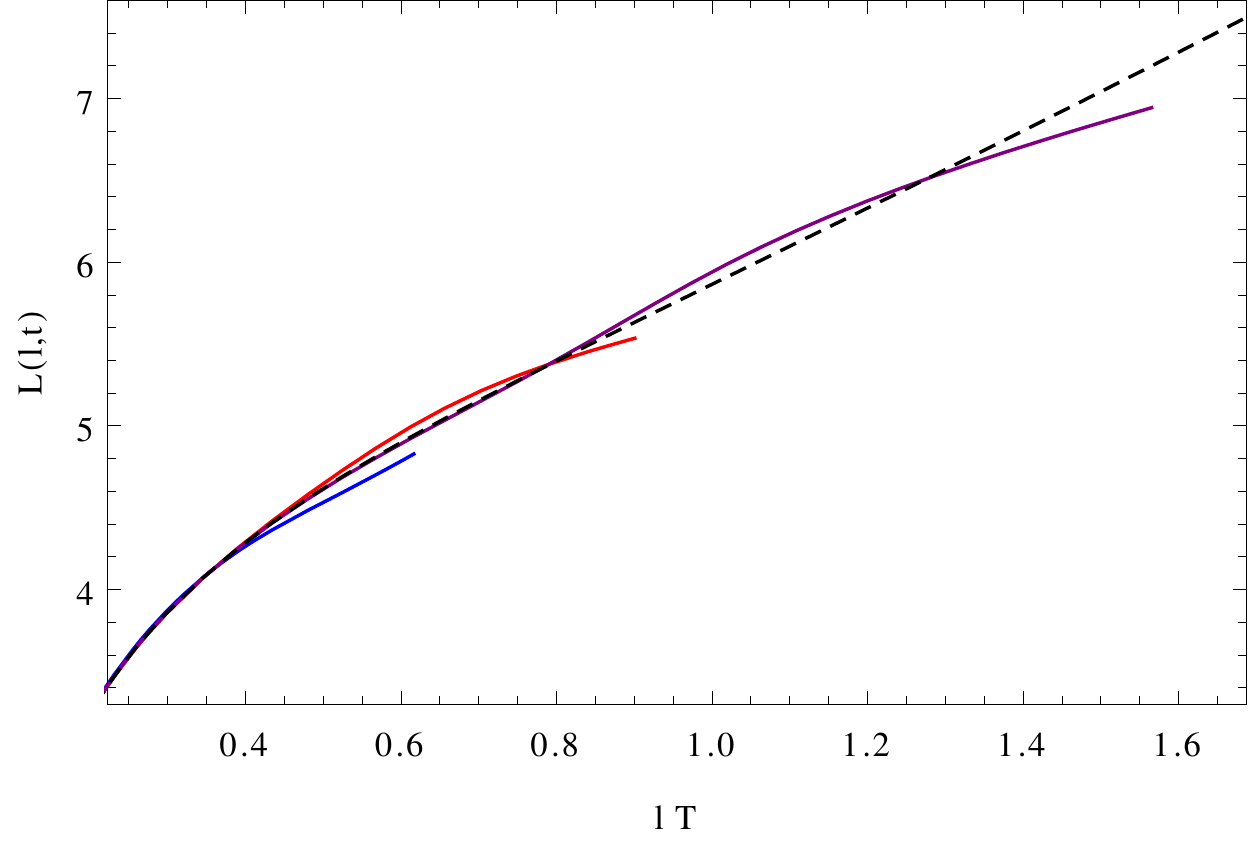}$\;\;\;\;\;$\includegraphics[scale=.6]{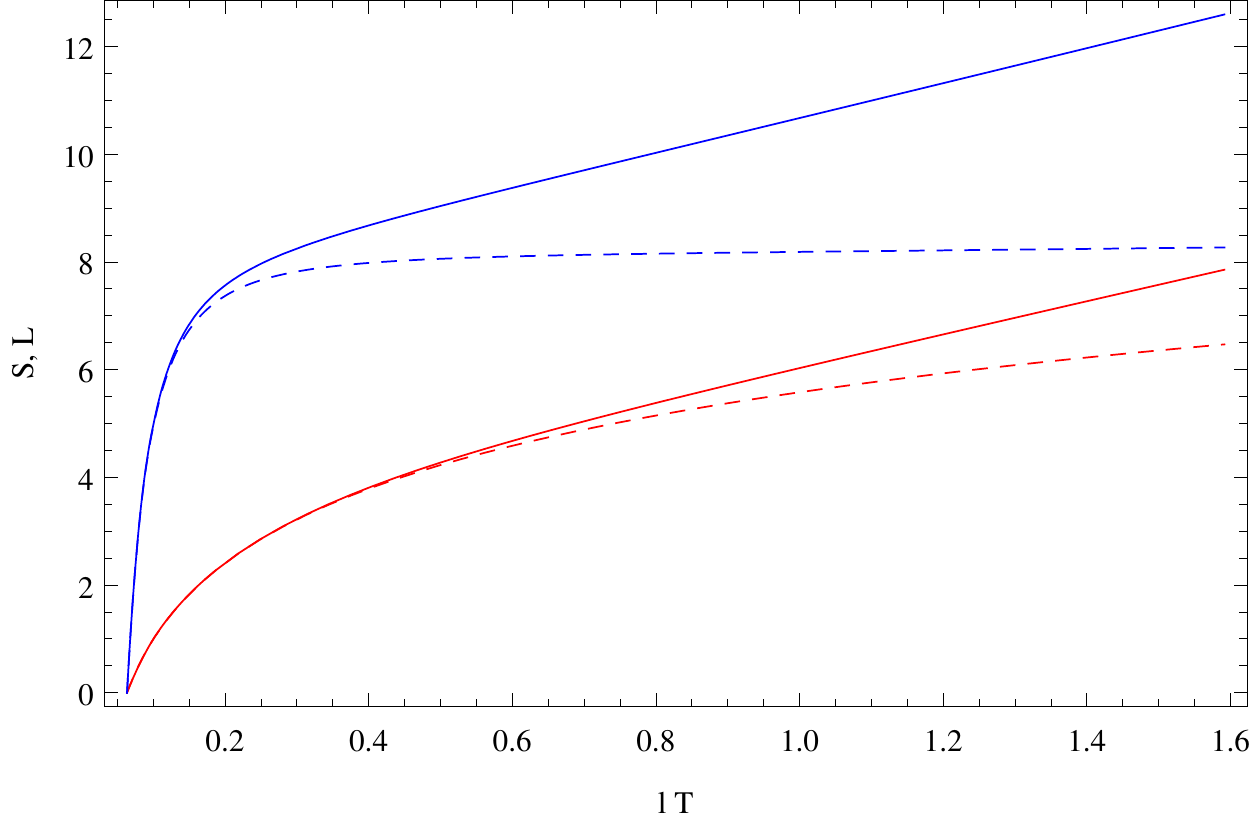}
\caption[Late time correlators and HEE as function of separation.]{\label{fixedT} 
Left: Renormalized length of geodesics as a function of their boundary  separation in  transverse directions for different fixed boundary times $t=0.5,\;1,\;1.5$ (endpoints from left to right). The curves terminate  when the geodesics leave the computational domain. 
The black dashed line shows  the thermal limit. 
Right: HEE (blue) and geodesic  length  (red) in the  thermal  (solid) and  zero temperature (dashed) limit.  
 }
\end{center}
\end{figure}

To conclude this section we finally study the departure of the length of the geodesics and HEE from equilibrium 
 for different times as a function of the boundary separation. This time we normalized the length of geodesics by subtracting a  cutoff dependent  piece.
The case for the 2-point function with separation in the transverse direction is  displayed in Fig.~\ref{fixedT}. 
Out of equilibrium effects manifest themselves as oscillations around the thermal value.  
The curves terminate when the geodesics leave the computational domain, so for early times we only have access to rather small boundary separations. The same effect is seen for HEE.

In the thermal limit the scaling for the  geodesic length   at small and  large  boundary separation is dictated by conformal symmetry and is proportional to
$2\log(l/2)$ and $2 l$ respectively.   
At large separation the HEE also scales linearly with the separation length, whereas  at small separation it is proportional to $1/l^2$. All these results agree precisely with the perturbative expressions derived in the limits of small and large temperatures \cite{Ryu:2006ef, Fischler:2012ca}. 
Our numerical results are shown in Fig.~\ref{fixedT} where we also plot the corresponding zero temperature results, which coincide with the thermal curves for small separations.

\section{Conclusions}\label{se:5}

In the paper at hand we have studied the time evolution of 2-point correlation functions in the geodesic approximation and holographic entanglement entropy in an anisotropic $\mathcal{N}=4$ SYM plasma.

To obtain the 2-point correlation function for separation in transverse and longitudinal direction we solved for the length of the geodesics in the anisotropic background \eqref{metric} by using a relaxation method with the zero temperature geodesic as an initial guess. At the $n$-th step the solution of the $(n-1)$-st step is used as initial guess. 

Choosing an infinite strip with finite separation either in the transverse or longitudinal direction and using the symmetries of the system the problem of finding minimal surfaces was reduced to finding geodesics in an auxiliary conformally related spacetime and the same strategy as for the 2-point correlation function goes through. 

The 2-point correlation functions  as well as holographic entanglement entropy show oscillatory behaviour around their thermal value, where  transverse and longitudinal directions oscillate out of phase.  
Both quantities approach their equilibrium value by exponentially damped oscillations and after the  initial far from equilibrium epoch the thermalization process is accurately  described by the lowest quasinormal mode.
That the late time behaviour of holographic entanglement entropy is captured by the ring down of the black brane geometry is one of the main results of this work and to our knowledge the first explicit verification of the connection between QNMs and holographic entanglement entropy found in \cite{Bhattacharya:2013bna}.

The methods developed and applied in this work can be generalized to other time-dependent asymptotically anti-de~Sitter backgrounds of interest, such as colliding shockwave backgrounds \cite{Grumiller:2008va,Gubser:2008pc}, which have been constructed numerically in the past few years \cite{Chesler:2010bi,Wu:2011yd,Casalderrey-Solana:2013aba}. Another interesting generalization would be the consideration of compact entangling regions, e.g.~with spherical topology to check the (in-)dependence of our results on the form of the entangling region.

\section*{Acknowledgments}

We thank Max Attems, Arjun Bagchi, Rudranil Basu, Paul Chesler, Jan de Boer, Ville Ker\"anen, Esperanza Lopez, Masahiro Nozaki, Florian Preis, Max Riegler, Paul Romatschke and Wilke van der Schee for discussions.

This work was supported by the following projects of the Austrian Science Fund (FWF): Y435-N16, I952-N16, P27182-N27, DKW1252-N27 and P26328.


\begin{appendix}

\section{Spectral method}\label{se:spectral}

In this appendix we give some details on how we numerically solve the Einstein equations (\ref{einstein}).
We work with the  inverse radial coordinate $z \equiv 1/r$ such that the boundary is located at $z=0$.
Due to asymptotic AdS boundary conditions, the metric functions $A$ and $\Sigma$ diverge as $z\to 0$.
It is numerically favourable to define new functions with the known divergent pieces removed and rescale them with appropriate powers of $z$ so that the resulting functions are finite or vanish as $z\to 0$; the precise boundary conditions on the new functions are presented in \eqref{BC} below.
This leads us to the following field redefinitions
\bse
\ba
A(z,v)&\to\frac{1}{z^2}+z A(z,v)\qquad B(z,v)& \to z^3 B(z,v)\\
\Sigma(z,v)&\to \frac{1}{z}+z^2\Sigma(z,v)\qquad \dot{\Sigma}(z,v)&\to\frac{1}{2z^2}+\frac{1}{2}z^2\dot{\Sigma}(z,v)\\
\ddot{\Sigma}(z,v)&\to\frac{1}{2z^3}+\frac{1}{4}\ddot{\Sigma}(z,v)\qquad \dot{B}(z,v)&\to-2z^3\dot{B}(z,v)\,.
\ea
\ese
The redefined anisotropy function $B$ allows to extract $b_4(t)$ simply from the boundary value of $B'$ 
\be
b_4(t)=B'(z=0,v=t),
\ee
where $B'=\partial_z B$.
In terms of the redefined fields the first four Einstein equations \eqref{einstein} can be rewritten in the form
\bse\label{redefEom}
\ba
\Sigma''+\frac{6}{z}\Sigma'+\left(\frac{6}{z^2}+\frac{9}{2}z^4B^2+3z^5BB'+\frac{1}{2}z^6B'^2\right)\Sigma&=&j_\Sigma\label{redefEom1}\\
\dot{\Sigma}'+\frac{2z^2(3\Sigma+z\Sigma')}{1+z^3\Sigma}\dot{\Sigma}&=&j_{\dot{\Sigma}}\label{redefEom2}\\
\dot{B}'+\frac{3(1+4z^3\Sigma+z^4\Sigma')}{2(z+z^4\Sigma)}\dot{B}&=&j_{\dot{B}}\label{redefEom3}\\
A''+\frac{4}{z}A'+\frac{2}{z^2}A&=&j_A\label{redefEom4}
\ea
\ese
with the source functions given by
\bse\label{source}
\ba
j_\Sigma&=&-\frac{9}{2}zB^2-3z^2BB'-\frac{1}{2}z^3B'^2\label{source1}\\
j_{\dot{\Sigma}}&=&-\frac{2(5\Sigma+2z^3\Sigma^2+z\Sigma')}{z^2+z^5\Sigma}\label{source2}\\
j_{\dot{B}}&=&\frac{3(1+z^4\dot{\Sigma})(3B+zB')}{8z^2(1+z^3\Sigma)}\label{source3}\\
j_A&=&-\frac{6(\Sigma(4+z^3\Sigma)+z^4\dot{B}(1+z^3\Sigma)^2(3B+zB')+z\Sigma'-z\dot{\Sigma}(1-2z^3\Sigma-z^4\Sigma'))}{(z+z^4\Sigma)^2}\,.\qquad\quad \label{source4}
\ea
\ese
The relation between dot-derivative and time-derivative, originally given in (\ref{dot}), turns into
\be\label{redefDot}
\dot{B}=-\frac{1}{2}\partial_v B+\frac{1}{4}B'+\frac{3}{4z}B+\frac{3}{4}z^2AB+\frac{1}{4}z^3B'A\,.
\ee
The boundary conditions for the redefined fields read
\bse\label{BC}
\ba
\Sigma(z=0,v)&=&0\qquad \Sigma'(z=0,v)=0\label{BC1}\\
\dot{\Sigma}(z=0,v)&=&a_4\label{BC2}\\
\dot{B}(z=0,v)&=&B'(z=0,v)\label{BC3}\\
A(z=0,v)&=&0\qquad A'(z=0,v)=a_4\label{BC4}
\ea
\ese
where we set $a_4=-1$. On the initial time slice the boundary condition (\ref{BC3}) is computed from the initial conditions
\be\label{IC}
B(z,v_0)=\beta z \exp\Big[-\frac{(z-z_0)^2}{\omega^2}\Big]
\ee
where we set $\beta=6.6$, $z_0=1/4$ and $\omega=1$.

For a given set of initial and boundary data the system of equations (\ref{redefEom}) allows for the following solution strategy:
\begin{enumerate}
\item With the initial conditions (\ref{IC}) and the boundary conditions (\ref{BC1}) we first solve (\ref{redefEom1}) for $\Sigma$ on the initial time slice.
\item For given $\Sigma$ and the boundary condition (\ref{BC2}) we solve (\ref{redefEom2}) for $\dot{\Sigma}$. 
\item Having $B$, $\Sigma$ and $\dot{\Sigma}$ we next solve (\ref{redefEom3}) for $\dot{B}$ using the boundary condition (\ref{BC3}). 
\item With $B$, $\Sigma$, $\dot{\Sigma}$ and $\dot{B}$ and the boundary conditions in (\ref{BC4}) we solve (\ref{redefEom4}) for $A$.
\item Finally we integrate (\ref{redefDot}) to get the new $B$ on the next time slice and repeat the whole procedure all over again.
\end{enumerate}
For constant $v$ we solve each of the equations (\ref{redefEom}) with a pseudo-spectral method \cite{boyd2001chebyshev} where the $N+1$ grid points $u_i$ on an intervall $[a,b]$ are located at
\be
z_i=\frac{1}{2}\big((a+b)+(a-b)\,\mathrm{cos}(i\pi/N)\big) \qquad i=1,\ldots,N+1.
\ee
In our computations we choose $z_i\in[0,1.6]$ and $N=60$.
These points can be used to construct the entries of the spectral differentiation matrix $D_{ij}$ \cite{Trefethen:2000:SMM:343368}:
\ba
D_{00}&=&\frac{2N^2+1}{6}\qquad D_{NN}=-\frac{2N^2+1}{6}\\
D_{jj}&=&\frac{-z_j}{2(1-z_j^2)}\qquad j=1,\ldots,N-1\\
D_{ij}&=&\frac{c_i}{c_j}\frac{(-1)^{i+j}}{(z_i-z_j)}\qquad j\ne j \quad i,j=0,\ldots,N
\ea
where $c_0=c_N=2$ and otherwise $c_i=1$.
The derivative of a function $f(z_i)\equiv f_i$ on the spectral grid points is obtained by multiplication with this diffentiation matrix
\be
f'_i=D_{ij}f_j\,.
\ee
This allows to turn each of the equations (\ref{redefEom}) into a system of linear equations. 
For example the second equation in (\ref{redefEom}) translates to
\be
L_{ij}\dot{\Sigma}_j=(j_{\dot{\Sigma}})_i
\ee
where the matrix $L_{ij}$ is given by
\be
L_{ij}=D_{ij}+\mathrm{diag}\Big[{\frac{2z^2(3\Sigma+z\Sigma')}{1+z^3\Sigma}}\Big]_{ij}
\ee
and the source vector $(j_{\dot{\Sigma}})_i$  by
\be
(j_{\dot{\Sigma}})_i=\Big[-\frac{2(5\Sigma+2z^3\Sigma^2+z\Sigma')}{z^2+z^5\Sigma}\Big]_i\,.
\ee
The boundary condition $\dot{\Sigma}_1=a_4$ is implemented by setting $(j_{\dot{\Sigma}})_1=a_4$ and $L_{1j}=\delta_{1j}$.
The solution vector $\dot{\Sigma}_i$ is then obtained by multiplying the inverse of $L_{ij}$ with the source vector
\be
\dot{\Sigma}_i=[L_{ij}]^{-1}(j_{\dot{\Sigma}})_j\,.
\ee 
We do this with the OCTAVE routine \texttt{linsolve}. The equations for $\Sigma_i$, $A_i$ and $\dot{B}_i$ can be solved in the same way.

To advance the solution for $B$ to the next time slice it  is sufficient to  use a simple fourth order Runge Kutta method \cite{Press:2007:NRE:1403886} 
\begin{equation}
B(z,v+\delta v)=B(z,v)+\delta v\, \Big(\tfrac{1}{6}k_1+\tfrac{1}{3}k_2+\tfrac{1}{3}k_3+\tfrac{1}{6}k_4\Big).
\end{equation}
where the koefficients $k_i$ are given by
\ba
k_1&=& \partial_v B(z,v)\\
k_2&=& \partial_v(B(z,v)+\tfrac{1}{2}k_1)\\
k_3&=&\partial_v(B(z,v)+\tfrac{1}{2}k_2)\\
k_4&=&\partial_v(B(z,v)+k_3)
\ea
and $\partial_v B$ is computed from \eqref{redefDot}.
In our simulations we use a stepsize of $\delta v\approx 0.01$ which is sufficient to achieve a stable time evolution.

\section{Relaxation method}\label{se:relax}

In relaxation methods differential equations are replaced by approximate finite difference equations (FDEs) on a discrete set of points. The solution is determined by starting with an inital guess and improving it iteratively. 
In this iterative procedure the result is said to relax to the true solution. 

First we define a grid $\sigma_i=h\,i$ of equidistant spacing $h=\tfrac{\sigma_N-\sigma_1}{N}$ with $i=1,\ldots,N$. We use a grid with $N=500$ points in all our simulations. The upper and lower bound of this grid are given by $\sigma_1\!=\!-1\!+\!\epsilon$ and $\sigma_N\!=\!1\!-\!\epsilon$ respectively, where $\epsilon$ denotes a UV cutoff. The discretized version of our trial solution on this grid is written  as
\be
X^\mu(\sigma_i)\equiv X^\mu_i=(V_i,Z_i,X_i)\,.
\ee
The explicit form of the trial solutions  used in the  computations of  2-point functions and  HEE are given in appendix \ref{app:B.1} and appendix \ref{app:B.2}, respectively.
In all our simulations we evaluate the cutoff $\epsilon$ such that the $z$-position of the corresponding cutoff surface is fixed at $Z_{UV}=0.05$, i.e.~at $r_{UV}=20$.
The boundary conditions are imposed at this cutoff surface
\ba
V_1&=&V_N=t\\
Z_1&=&Z_N=Z_{UV}\\
X_1&=&-l/2\qquad X_N=l/2
\ea
where $t$ denotes the time and $l$ the separation on the cutoff surface. 

The finite difference representation of the geodesic equation (\ref{geodesic2}) is given by
\bse\label{FDE}
\ba
E_i^1&=& V_{i+1} - V_{i} -h (\bar{p}_V)_{i}\\
E_i^2&=& Z_{i+1} - Z_{i} -h (\bar{p}_X)_{i}\\
E_i^3&=& X_{i+1} - X_{i} -h (\bar{p}_X)_{i}\\
E_i^4&=& ({p}_V)_{i+1}-({p}_V)_{i}+h\bar{J}_i (\bar{p}_V)_{i}+h((\bar{\Gamma}^V{}_{VV})_i(\bar{p}_V)_{i}^2+(\bar{\Gamma}^V{}_{XX})_i(\bar{p}_X)_{i}^2)\\
E_i^5&=& ({p}_Z)_{i+1}-({p}_Z)_{i}+h\bar{J}_i (\bar{p}_Z)_{i}+h((\bar{\Gamma}^Z{}_{VV})_i(\bar{p}_V)_{i}^2\nonumber\\&& \quad+2(\bar{\Gamma}^Z{}_{VZ})_i(\bar{p}_V)_{i}(\bar{p}_Z)_{i}+\bar{\Gamma}^Z{}_{ZZ})_i(\bar{p}_Z)_{i}^2+\bar{\Gamma}^Z{}_{XX})_i(\bar{p}_X)_{i}^2)\\
E_i^6&=& ({p}_X)_{i+1}-({p}_X)_{i}+h\bar{J}_i (\bar{p}_X)_{i}+2h((\bar{\Gamma}^X{}_{XV})_i(\bar{p}_X)_{i}(\bar{p}_V)_{i}+(\bar{\Gamma}^X{}_{XZ})_i(\bar{p}_X)_{i}(\bar{p}_Z)_{i})\qquad
\ea
\ese
where $E_i^k$ is the residual at point $i$ in equation $k$; further $(p_V)_i=\dot{V}_i$ denotes the first derivatives of the geodesic coordinates; quantitities with bar  are averaged like $(\bar{p}_V)_i=\frac{(p_V)_{i-1}+(p_V)_i}{2}$; the Christoffel symbols  $(\bar{\Gamma}^X{}_{XZ})_i$ are evaluated from the averaged metric functions;
the explicit form of the Jacobian $\bar{J}_i$ for the case of  2-point functions and HEE  are given in appendix \ref{app:B.1} and appendix \ref{app:B.2}.

The initial guess will in general not satisfy these FDEs very well, i.e.\! the residua $E^k_i$ will be rather large. To quantify the deviation of a given trial solution to the true solution we use the following measure
\be
\delta=\frac{\sum_{i,k} |E_i^k|}{6N}\,.
\ee
The strategy is to compute increments $\Delta X^\mu$ for the geodesic coordinates $X^\mu$ such that $X^\mu_{\rm new}=X_{\rm old}^\mu+\Delta X^\mu$ is an improved approximation to the previous solution $X_{\rm old}^\mu$. This we do iteratively until we reach 
\be
\delta<10^{-15}\,.
\label{eq:lalapetz}
\ee
Equations for the increments are obtained by demanding the first order Taylor expansion of the finite difference equations with respect to small changes in the coordinates to vanish. We do this following exactly the procedure described in the section on relaxation methods in reference \cite{Press:2007:NRE:1403886}.

The correction $\Delta X^\mu$ generated  from the first order Taylor expansion is in general only an improvement close to the true solution. We account for this by introducing  a weight $\alpha$ that modifies the correction in each relaxation step 
\be
X^\mu_{\rm new}=X^\mu_{\rm old}+\alpha\Delta X^\mu ,
\ee
We choose the weight $\alpha$ such that the full correction is used only close to the true solution. 
\be 
\alpha =
  \begin{cases}
    0.05 & \quad \text{if } \delta \ge 0.05\\
    1    & \quad \text{else } \\
  \end{cases}
\ee

In the time evolution we use as ansatz geodesic at time $t_{i}=t_{i-1}+\delta t$ the geodesic form the previous time step $t_{i-1}$.
With a step size of $\delta t \approx 0.05$ usually less than 5 relaxation steps turned out to be sufficient to reach our accuracy goal of $\delta < 10^{-15}$.

\subsection{Initialization with Poincar\'e patch AdS geodesics}\label{app:B.1}

The line element in 3-dimensional Poincar\'e patch AdS space is given by
\be
\extd s^2=\frac{1}{z^2}\(-\extd v^2-2 \extd z \extd v + \extd x^2\)\,.
\ee
We parametrize the geodesics by an affine parameter $\tau$ and denote derivatives with respect to it by dot. The geodesic equations of motion allow first integrals
\be
\dot{x}=L z^2\qquad \dot{v}=Ez^2 -\dot{z}\qquad \dot{z}=\pm z\, \sqrt{1-(L^2-E^2)z^2}
\ee
where $L$ and $E$ are constants of motion and the third equation comes from the spacelike condition $\extd s^2=1$. The last equation shows that the geodesics have two branches. It is convenient to reparametrize both branches of the geodesics in terms of the holographic coordinate $z$, i.e., to find expressions $x_\pm(z)$ and $v_\pm(z)$. We want to initialize the relaxation code with geodesics that have a symmetric advanced time with respect to the exchange of the branches, $v_+(z)=v_-(z)$. This is achieved by setting $E=0$. Then integrating the above system yields 
\be\label{firstIntegral}
x_\pm(z)=\pm \sqrt{L^{-2} - z^2}\; \qquad v(z)=v_0-z
\ee
where $v_0$ is the boundary time and we have set to zero the additive integration constant in $x_\pm$. In the solution for the advanced time coordinate one clearly sees the `bending back in time'-effect as one goes deeper into the bulk. By this we mean simply that $v(z)$ becomes more negative as the holographic coordinate $z$ increases.

With the choices above the constant of motion $L$ is related to the separation $l$  of the two endpoints  on the boundary 
\be
l = \big|x_+-x_-\big|_{z=0} = \frac 2L\,.  
\ee
For our numerical algorithm it turns out to be useful to choose a non-affine parametrization that lies in a fixed interval and covers both branches at the same time, namely 
\ba\label{nonAffine}
z(\s)=\frac{l}{2}\big(1-\s^2\big)\qquad x(\s)=\frac{l}{2}\big(\s \sqrt{2-\s^2}\big)\qquad v(\s)=v_0-z(\s)
\ea
where $\s \in [-1,1]$.
We realize a UV-cutoff at a given value $Z_{UV}$ by truncating the non-affine parater $\sigma\in[-1\!+\!\epsilon,1\!-\!\epsilon]$ with $\epsilon$ given by
\be
\epsilon=1-\sqrt{1-\frac{2 Z_{UV}}{l}}.
\ee
To get the Jacobian we need in the non-affine geodesic equation we first integrate the third equation in \eqref{firstIntegral} to express the affine parameter $\tau$ in terms of $z$
\be
\tau(z)=\pm\int \frac{\extd z}{z \sqrt{1-\tfrac{l^2}{4}z^2}}=\mp \mathrm{artanh}\Big(\sqrt{1-\tfrac{l^2}{4}z^2}\Big).
\ee
Next we use the first equation in \eqref{nonAffine} and express the affine parameter $\tau$ in terms of the non-affine parameter $\sigma$
\be
\tau(\sigma)=\mp \mathrm{artanh}\Big(\sigma \sqrt{2-\sigma^2}\Big)\,.
\ee
The Jacobian is then given by
\be
J(\sigma)=\frac{\extd^2\tau}{\extd\sigma^2}\Big/\frac{\extd\tau}{\extd\sigma}=\frac{5\sigma-3\sigma^3}{2-3\sigma^2+\sigma^4}\,.
\ee

\subsection{Initialization with conformal AdS geodesics}\label{app:B.2}

We proceed here similarly to appendix \ref{app:B.1}. Starting with either of the line elements \eqref{eq:angelinajolie} in the isotropic static limit ($B=0$, $\Sigma=r$, $A=r^2$) and introducing again $z=1/r$ yields
\be
\extd s^2=\frac{1}{z^6}\(-\extd v^2-2 \extd z \extd v + \extd x^2\)
\ee
where we called the third coordinate $x$, regardless of the specific case.
Integrating the geodesic equations, using $\extd s^2=1$ and setting $E=0$ yields
\be\label{eoms}
\dot{x}=L z^6\qquad \dot{v}= -\dot{z}\qquad \dot{z} = \pm z^3\sqrt{1-L^2z^6}\,.
\ee
With help of the identity
\begin{align}
\int \frac{\extd z z^n}{\sqrt{1-L^2z^6}}=\frac{z^{1+n}}{1+n}\, {}_2F_1\lk \tfrac{1}{2},\tfrac{1+n}{6},1+\tfrac{1+n}{6};L^2z^6\rk\,
\end{align}
the third equation in (\ref{eoms}) can be used to express the affine parameter in terms of the holographic coordinate
\begin{align}
\tau=\pm \int \frac{\extd z}{z^3\sqrt{1-L^2z^6}}=\mp \frac{1}{2z^2}\, {}_2F_1\lk \tfrac{1}{2},-\tfrac{1}{3},\tfrac{2}{3};L^2z^6\rk\,.
\end{align}
In order to obtain the solution for the geodesic it turns out to be convenient to solve for $x$ in terms of $z$
\be\label{conformalgeo}
x_\pm=\pm\int \frac{\extd z \,L z^3}{\sqrt{1-L^2z^6}}=\mp \frac{l}{2} \pm \frac{L z^4}{4}\, {}_2F_1 \lk \tfrac{1}{2},\tfrac{2}{3},\tfrac{5}{3}; L^2 z^6\rk\,.
\ee
\begin{figure}
\begin{center}
\includegraphics[scale=0.5]{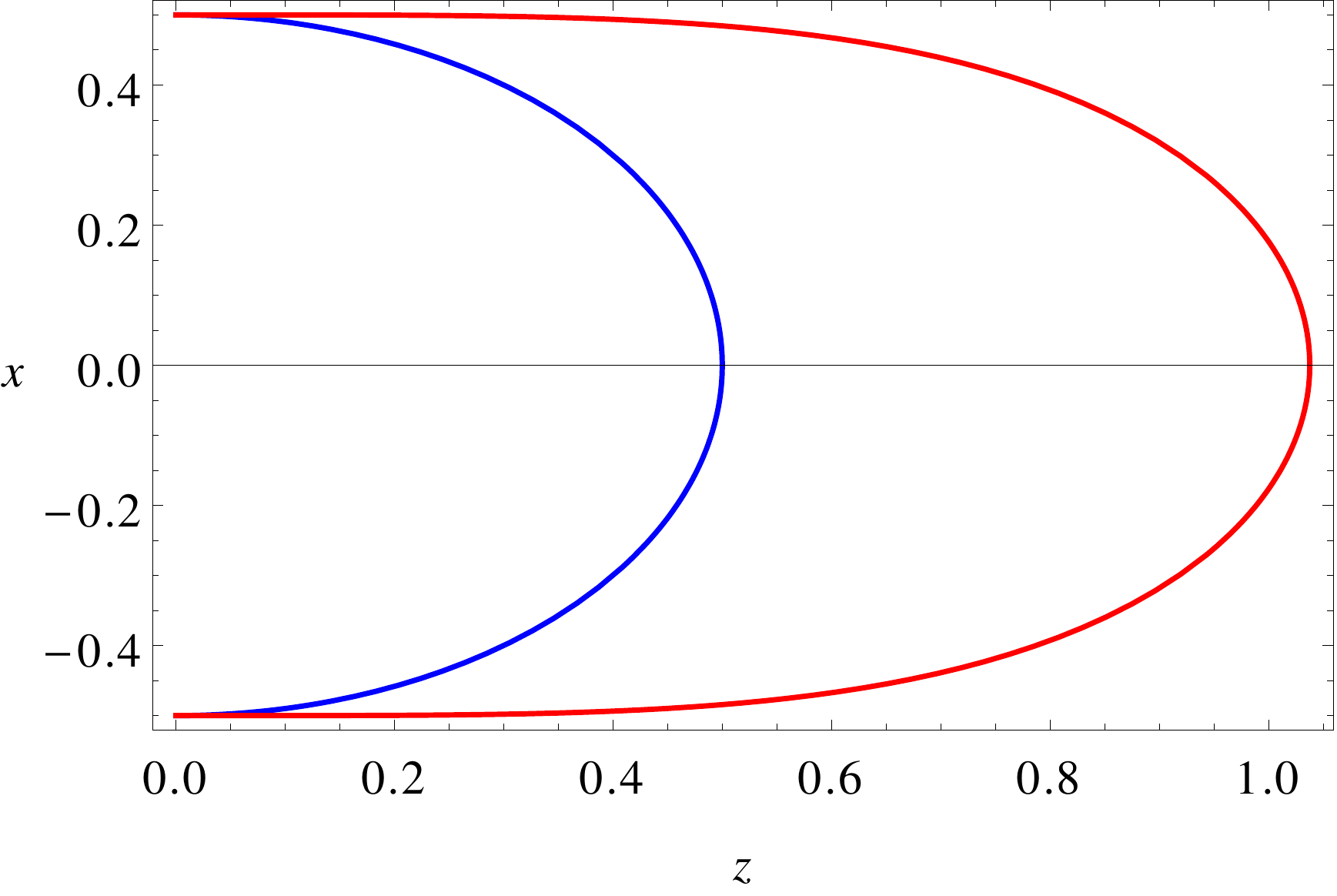}
\caption[Comparison between AdS and conformal geodesics.]{\label{confgeo} 
Comparison of the pure AdS geodesic (blue) with the conformal geodesic (red). For the same separation the conformal geodesic extends about twice as far into the bulk. 
 }
\end{center}
\end{figure}

The subscript $\pm$ corresponds to the two branches of the geodesic and $l/2$ is an integration constant corresponding to the boundary value of the geodesic since the hypergeometric function is zero at the boundary $z=0$.
The endpoint of the geodesic is given by $L^2z^6=1$, meaning that the two branches  do not join in general. 
In order to get a smoothly joining geodesic we need to adjust $L$ and express it in terms of the boundary separation $l$
\be
L=\frac{\pi^\frac{3}{2} \Gamma\lk 5/3\rk^3}{8 l^3 \Gamma\lk 7/6\rk^3}\,. 
\ee
Like for the Poincar\'e patch AdS geodesics we choose a non-affine parametrization
\ba\label{nonAffine2}
z(\s)&=&Z_{\rm max}\big(1-\s^2\big)\\
x(\s)&=&\mathrm{sgn}(\sigma)\Big(-\frac{l}{2} + \frac{L z(\sigma)^4}{4}\, {}_2F_1 \lk \tfrac{1}{2},\tfrac{2}{3},\tfrac{5}{3}; L^2 z(\sigma)^6\rk\Big)\\
v(\s)&=&v_0-z(\s)
\ea
where $\s \in [-1,1]$ and $Z_{\rm max}=\frac{2l}{\sqrt{\pi}}\Gamma(\tfrac{7}{6})\Big/\Gamma(\tfrac{5}{3})$ denotes the $z$-position at which the two branches join.
We realize a UV-cutoff at a given value $Z_{UV}$ by truncating the non-affine parater $\sigma\in[-1\!+\!\epsilon,1\!-\!\epsilon]$ with $\epsilon$ given by
\be
\epsilon=1-\sqrt{1-\frac{Z_{UV}}{Z_{\rm max}}}\,.
\ee
The affine parameter $\tau$ in terms of the non-affine parameter $\sigma$ reads
\be
\tau(\sigma)=\mp \frac{1}{2Z_{\rm max}^2(1-\sigma^2)}\, {}_2F_1\lk \tfrac{1}{2},-\tfrac{1}{3},\tfrac{2}{3};L^2 Z_{\rm max}^{12}(1-\sigma^2)^6\rk\,.
\ee
The Jacobian is then given by
\be
J(\sigma)=\frac{\extd^2\tau}{\extd\sigma^2}\Big/\frac{\extd\tau}{\extd\sigma}=\frac{-51\s+145\s^3-205\s^5+159\s^7-65\s^9+11\s^{11}}{(2-\s^2)(1-\s^2)(3-3\s^2+\s^4)(1-\s^2+\s^4)}\,.
\ee
As can be seen from Fig.~\ref{confgeo} for the same boundary separation the conformal geodesic extends much further into the bulk compared to the Poincar\'e patch AdS case. As a consequence, the boundary separation we can choose for the HEE is much smaller than for the 2-point functions for any given value of the boundary time.

\end{appendix}



\begin{thebibliography}{10}
\addcontentsline{toc}{section}{References}

\bibitem{Teaney:2000cw}
D.~Teaney, J.~Lauret and E.~V. Shuryak, {\it {Flow at the SPS and RHIC as a
  quark gluon plasma signature}},  {\em Phys.Rev.Lett.} {\bf 86} (2001)
  4783--4786 [\href{http://arXiv.org/abs/nucl-th/0011058}{{\tt
  nucl-th/0011058}}].

\bibitem{Huovinen:2001cy}
P.~Huovinen, P.~Kolb, U.~W. Heinz, P.~Ruuskanen and S.~Voloshin, {\it {Radial
  and elliptic flow at RHIC: Further predictions}},  {\em Phys.Lett.} {\bf
  B503} (2001) 58--64 [\href{http://arXiv.org/abs/hep-ph/0101136}{{\tt
  hep-ph/0101136}}].

\bibitem{Hirano:2002ds}
T.~Hirano and K.~Tsuda, {\it {Collective flow and two pion correlations from a
  relativistic hydrodynamic model with early chemical freezeout}},  {\em
  Phys.Rev.} {\bf C66} (2002) 054905
  [\href{http://arXiv.org/abs/nucl-th/0205043}{{\tt nucl-th/0205043}}].

\bibitem{Romatschke:2007mq}
P.~Romatschke and U.~Romatschke, {\it {Viscosity Information from Relativistic
  Nuclear Collisions: How Perfect is the Fluid Observed at RHIC?}},  {\em
  Phys.Rev.Lett.} {\bf 99} (2007) 172301
  [\href{http://arXiv.org/abs/0706.1522}{{\tt 0706.1522}}].

\bibitem{Polkovnikov:2010yn}
A.~Polkovnikov, K.~Sengupta, A.~Silva and M.~Vengalattore, {\it {Nonequilibrium
  dynamics of closed interacting quantum systems}},  {\em Rev.Mod.Phys.} {\bf
  83} (2011) 863 [\href{http://arXiv.org/abs/1007.5331}{{\tt 1007.5331}}].

\bibitem{Maldacena:1997re}
J.~M. Maldacena, {\it {The Large N limit of superconformal field theories and
  supergravity}},  {\em Int.J.Theor.Phys.} {\bf 38} (1999) 1113--1133
  [\href{http://arXiv.org/abs/hep-th/9711200}{{\tt hep-th/9711200}}].

\bibitem{Witten:1998qj}
E.~Witten, {\it {Anti-de Sitter space and holography}},  {\em Adv. Theor. Math.
  Phys.} {\bf 2} (1998) 253--291
  [\href{http://arXiv.org/abs/hep-th/9802150}{{\tt hep-th/9802150}}].

\bibitem{Gubser:1998bc}
S.~Gubser, I.~R. Klebanov and A.~M. Polyakov, {\it {Gauge theory correlators
  from noncritical string theory}},  {\em Phys.Lett.} {\bf B428} (1998)
  105--114 [\href{http://arXiv.org/abs/hep-th/9802109}{{\tt hep-th/9802109}}].

\bibitem{Witten:1998zw}
E.~Witten, {\it {Anti-de Sitter space, thermal phase transition, and
  confinement in gauge theories}},  {\em Adv. Theor. Math. Phys.} {\bf 2}
  (1998) 505--532 [\href{http://arXiv.org/abs/hep-th/9803131}{{\tt
  hep-th/9803131}}].

\bibitem{Eisert:2008ur}
J.~Eisert, M.~Cramer and M.~Plenio, {\it {Area laws for the entanglement
  entropy - a review}},  {\em Rev.Mod.Phys.} {\bf 82} (2010) 277--306
  [\href{http://arXiv.org/abs/0808.3773}{{\tt 0808.3773}}].

\bibitem{Ryu:2006bv}
S.~Ryu and T.~Takayanagi, {\it {Holographic derivation of entanglement entropy
  from AdS/CFT}},  {\em Phys.Rev.Lett.} {\bf 96} (2006) 181602
  [\href{http://arXiv.org/abs/hep-th/0603001}{{\tt hep-th/0603001}}].

\bibitem{Nishioka:2009un}
T.~Nishioka, S.~Ryu and T.~Takayanagi, {\it {Holographic Entanglement Entropy:
  An Overview}},  {\em J.Phys.} {\bf A42} (2009) 504008
  [\href{http://arXiv.org/abs/0905.0932}{{\tt 0905.0932}}].

\bibitem{Hubeny:2007xt}
V.~E. Hubeny, M.~Rangamani and T.~Takayanagi, {\it {A Covariant holographic
  entanglement entropy proposal}},  {\em JHEP} {\bf 0707} (2007) 062
  [\href{http://arXiv.org/abs/0705.0016}{{\tt 0705.0016}}].

\bibitem{Calabrese:2004eu}
P.~Calabrese and J.~L. Cardy, {\it {Entanglement entropy and quantum field
  theory}},  {\em J.Stat.Mech.} {\bf 0406} (2004) P06002
  [\href{http://arXiv.org/abs/hep-th/0405152}{{\tt hep-th/0405152}}].

\bibitem{Holzhey:1994we}
C.~Holzhey, F.~Larsen and F.~Wilczek, {\it {Geometric and renormalized entropy
  in conformal field theory}},  {\em Nucl.Phys.} {\bf B424} (1994) 443--467
  [\href{http://arXiv.org/abs/hep-th/9403108}{{\tt hep-th/9403108}}].

\bibitem{Vidal:2002rm}
G.~Vidal, J.~Latorre, E.~Rico and A.~Kitaev, {\it {Entanglement in quantum
  critical phenomena}},  {\em Phys.Rev.Lett.} {\bf 90} (2003) 227902
  [\href{http://arXiv.org/abs/quant-ph/0211074}{{\tt quant-ph/0211074}}].

\bibitem{Bagchi:2014iea}
A.~Bagchi, R.~Basu, D.~Grumiller and M.~Riegler, {\it {Entanglement entropy in
  Galilean conformal field theories and flat holography}},  {\em
  Phys.Rev.Lett.} {\bf 114} (2015), no.~11 111602
  [\href{http://arXiv.org/abs/1410.4089}{{\tt 1410.4089}}].

\bibitem{Calabrese:2005in}
P.~Calabrese and J.~L. Cardy, {\it {Evolution of entanglement entropy in
  one-dimensional systems}},  {\em J.Stat.Mech.} {\bf 0504} (2005) P04010
  [\href{http://arXiv.org/abs/cond-mat/0503393}{{\tt cond-mat/0503393}}].

\bibitem{AbajoArrastia:2010yt}
J.~Abajo-Arrastia, J.~Aparicio and E.~Lopez, {\it {Holographic Evolution of
  Entanglement Entropy}},  {\em JHEP} {\bf 1011} (2010) 149
  [\href{http://arXiv.org/abs/1006.4090}{{\tt 1006.4090}}].

\bibitem{Balasubramanian:2011ur}
V.~Balasubramanian, A.~Bernamonti, J.~de~Boer, N.~Copland, B.~Craps {\em
  et.~al.}, {\it {Holographic Thermalization}},  {\em Phys.Rev.} {\bf D84}
  (2011) 026010 [\href{http://arXiv.org/abs/1103.2683}{{\tt 1103.2683}}].

\bibitem{Albash:2010mv}
T.~Albash and C.~V. Johnson, {\it {Evolution of Holographic Entanglement
  Entropy after Thermal and Electromagnetic Quenches}},  {\em New J.Phys.} {\bf
  13} (2011) 045017 [\href{http://arXiv.org/abs/1008.3027}{{\tt 1008.3027}}].

\bibitem{Baron:2012fv}
W.~Baron, D.~Galante and M.~Schvellinger, {\it {Dynamics of holographic
  thermalization}},  {\em JHEP} {\bf 1303} (2013) 070
  [\href{http://arXiv.org/abs/1212.5234}{{\tt 1212.5234}}].

\bibitem{Galante:2012pv}
D.~Galante and M.~Schvellinger, {\it {Thermalization with a chemical potential
  from AdS spaces}},  {\em JHEP} {\bf 1207} (2012) 096
  [\href{http://arXiv.org/abs/1205.1548}{{\tt 1205.1548}}].

\bibitem{Keranen:2011xs}
V.~Keranen, E.~Keski-Vakkuri and L.~Thorlacius, {\it {Thermalization and
  entanglement following a non-relativistic holographic quench}},  {\em
  Phys.Rev.} {\bf D85} (2012) 026005
  [\href{http://arXiv.org/abs/1110.5035}{{\tt 1110.5035}}].

\bibitem{Liu:2013qca}
H.~Liu and S.~J. Suh, {\it {Entanglement growth during thermalization in
  holographic systems}},  {\em Phys.Rev.} {\bf D89} (2014), no.~6 066012
  [\href{http://arXiv.org/abs/1311.1200}{{\tt 1311.1200}}].

\bibitem{Liu:2013iza}
H.~Liu and S.~J. Suh, {\it {Entanglement Tsunami: Universal Scaling in
  Holographic Thermalization}},  {\em Phys.Rev.Lett.} {\bf 112} (2014) 011601
  [\href{http://arXiv.org/abs/1305.7244}{{\tt 1305.7244}}].

\bibitem{Keranen:2014zoa}
V.~Keranen, H.~Nishimura, S.~Stricker, O.~Taanila and A.~Vuorinen, {\it
  {Dynamics of gravitational collapse and holographic entropy production}},
  {\em Phys.Rev.} {\bf D90} (2014), no.~6 064033
  [\href{http://arXiv.org/abs/1405.7015}{{\tt 1405.7015}}].

\bibitem{Keranen:2015fqa}
V.~Keranen, H.~Nishimura, S.~Stricker, O.~Taanila and A.~Vuorinen, {\it
  {Gravitational collapse of thin shells: Time evolution of the holographic
  entanglement entropy}},  \href{http://arXiv.org/abs/1502.01277}{{\tt
  1502.01277}}.

\bibitem{Alishahiha:2014cwa}
M.~Alishahiha, A.~F. Astaneh and M.~R.~M. Mozaffar, {\it {Thermalization in
  backgrounds with hyperscaling violating factor}},  {\em Phys.Rev.} {\bf D90}
  (2014), no.~4 046004 [\href{http://arXiv.org/abs/1401.2807}{{\tt
  1401.2807}}].

\bibitem{Fonda:2014ula}
P.~Fonda, L.~Franti, V.~Ker{\"a}nen, E.~Keski-Vakkuri, L.~Thorlacius {\em
  et.~al.}, {\it {Holographic thermalization with Lifshitz scaling and
  hyperscaling violation}},  {\em JHEP} {\bf 1408} (2014) 051
  [\href{http://arXiv.org/abs/1401.6088}{{\tt 1401.6088}}].

\bibitem{Bizon:2011gg}
P.~Bizon and A.~Rostworowski, {\it {On weakly turbulent instability of anti-de
  Sitter space}},  {\em Phys.Rev.Lett.} {\bf 107} (2011) 031102
  [\href{http://arXiv.org/abs/1104.3702}{{\tt 1104.3702}}].

\bibitem{Abajo-Arrastia:2014fma}
J.~Abajo-Arrastia, E.~da~Silva, E.~Lopez, J.~Mas and A.~Serantes, {\it
  {Holographic Relaxation of Finite Size Isolated Quantum Systems}},  {\em
  JHEP} {\bf 1405} (2014) 126 [\href{http://arXiv.org/abs/1403.2632}{{\tt
  1403.2632}}].

\bibitem{Buchel:2014gta}
A.~Buchel, R.~C. Myers and A.~van Niekerk, {\it {Nonlocal probes of
  thermalization in holographic quenches with spectral methods}},  {\em JHEP}
  {\bf 1502} (2015) 017 [\href{http://arXiv.org/abs/1410.6201}{{\tt
  1410.6201}}].

\bibitem{daSilva:2014zva}
E.~da~Silva, E.~Lopez, J.~Mas and A.~Serantes, {\it {Collapse and Revival in
  Holographic Quenches}},  {\em JHEP} {\bf 1504} (2015) 038
  [\href{http://arXiv.org/abs/1412.6002}{{\tt 1412.6002}}].

\bibitem{Muller:2011ra}
B.~Muller and A.~Schafer, {\it {Entropy Creation in Relativistic Heavy Ion
  Collisions}},  {\em Int.J.Mod.Phys.} {\bf E20} (2011) 2235--2267
  [\href{http://arXiv.org/abs/1110.2378}{{\tt 1110.2378}}].

\bibitem{Narayan:2012ks}
K.~Narayan, T.~Takayanagi and S.~P. Trivedi, {\it {AdS plane waves and
  entanglement entropy}},  {\em JHEP} {\bf 1304} (2013) 051
  [\href{http://arXiv.org/abs/1212.4328}{{\tt 1212.4328}}].

\bibitem{Chesler:2008hg}
P.~M. Chesler and L.~G. Yaffe, {\it {Horizon formation and far-from-equilibrium
  isotropization in supersymmetric Yang-Mills plasma}},  {\em Phys.Rev.Lett.}
  {\bf 102} (2009) 211601 [\href{http://arXiv.org/abs/0812.2053}{{\tt
  0812.2053}}].

\bibitem{Chesler:2009cy}
P.~M. Chesler and L.~G. Yaffe, {\it {Boost invariant flow, black hole
  formation, and far-from-equilibrium dynamics in N = 4 supersymmetric
  Yang-Mills theory}},  {\em Phys.Rev.} {\bf D82} (2010) 026006
  [\href{http://arXiv.org/abs/0906.4426}{{\tt 0906.4426}}].

\bibitem{Heller:2012km}
M.~P. Heller, D.~Mateos, W.~van~der Schee and D.~Trancanelli, {\it {Strong
  Coupling Isotropization of Non-Abelian Plasmas Simplified}},  {\em
  Phys.Rev.Lett.} {\bf 108} (2012) 191601
  [\href{http://arXiv.org/abs/1202.0981}{{\tt 1202.0981}}].

\bibitem{Heller:2013oxa}
M.~P. Heller, D.~Mateos, W.~van~der Schee and M.~Triana, {\it {Holographic
  isotropization linearized}},  {\em JHEP} {\bf 1309} (2013) 026
  [\href{http://arXiv.org/abs/1304.5172}{{\tt 1304.5172}}].

\bibitem{Chesler:2013lia}
P.~M. Chesler and L.~G. Yaffe, {\it {Numerical solution of gravitational
  dynamics in asymptotically anti-de Sitter spacetimes}},  {\em JHEP} {\bf
  1407} (2014) 086 [\href{http://arXiv.org/abs/1309.1439}{{\tt 1309.1439}}].

\bibitem{Henningson:1998gx}
M.~Henningson and K.~Skenderis, {\it {The Holographic Weyl anomaly}},  {\em
  JHEP} {\bf 9807} (1998) 023 [\href{http://arXiv.org/abs/hep-th/9806087}{{\tt
  hep-th/9806087}}].

\bibitem{deHaro:2000xn}
S.~de~Haro, S.~N. Solodukhin and K.~Skenderis, {\it {Holographic reconstruction
  of space-time and renormalization in the AdS / CFT correspondence}},  {\em
  Commun.Math.Phys.} {\bf 217} (2001) 595--622
  [\href{http://arXiv.org/abs/hep-th/0002230}{{\tt hep-th/0002230}}].

\bibitem{Balasubramanian:1999zv}
V.~Balasubramanian and S.~F. Ross, {\it {Holographic particle detection}},
  {\em Phys.Rev.} {\bf D61} (2000) 044007
  [\href{http://arXiv.org/abs/hep-th/9906226}{{\tt hep-th/9906226}}].

\bibitem{Festuccia:2005pi}
G.~Festuccia and H.~Liu, {\it {Excursions beyond the horizon: Black hole
  singularities in Yang-Mills theories. I.}},  {\em JHEP} {\bf 0604} (2006) 044
  [\href{http://arXiv.org/abs/hep-th/0506202}{{\tt hep-th/0506202}}].

\bibitem{Keranen:2014lna}
V.~Keranen and P.~Kleinert, {\it {Non-equilibrium scalar two point functions in
  AdS/CFT}},  \href{http://arXiv.org/abs/1412.2806}{{\tt 1412.2806}}.

\bibitem{Ryu:2006ef}
S.~Ryu and T.~Takayanagi, {\it {Aspects of Holographic Entanglement Entropy}},
  {\em JHEP} {\bf 0608} (2006) 045
  [\href{http://arXiv.org/abs/hep-th/0605073}{{\tt hep-th/0605073}}].

\bibitem{octave:2014}
S.~H. John W.~Eaton, David~Bateman and R.~Wehbring, {\em {GNU Octave} version
  3.8.1 manual: a high-level interactive language for numerical computations}.
\newblock CreateSpace Independent Publishing Platform, 2014.
\newblock {ISBN} 1441413006.

\bibitem{Press:2007:NRE:1403886}
W.~H. Press, S.~A. Teukolsky, W.~T. Vetterling and B.~P. Flannery, {\em
  Numerical Recipes 3rd Edition: The Art of Scientific Computing}.
\newblock Cambridge University Press, New York, NY, USA, 3~ed., 2007.

\bibitem{Starinets:2002br}
A.~O. Starinets, {\it {Quasinormal modes of near extremal black branes}},  {\em
  Phys.Rev.} {\bf D66} (2002) 124013
  [\href{http://arXiv.org/abs/hep-th/0207133}{{\tt hep-th/0207133}}].

\bibitem{Kovtun:2005ev}
P.~K. Kovtun and A.~O. Starinets, {\it {Quasinormal modes and holography}},
  {\em Phys.Rev.} {\bf D72} (2005) 086009
  [\href{http://arXiv.org/abs/hep-th/0506184}{{\tt hep-th/0506184}}].

\bibitem{Birmingham:2001pj}
D.~Birmingham, I.~Sachs and S.~N. Solodukhin, {\it {Conformal field theory
  interpretation of black hole quasinormal modes}},  {\em Phys.Rev.Lett.} {\bf
  88} (2002) 151301 [\href{http://arXiv.org/abs/hep-th/0112055}{{\tt
  hep-th/0112055}}].

\bibitem{Son:2002sd}
D.~T. Son and A.~O. Starinets, {\it {Minkowski space correlators in AdS / CFT
  correspondence: Recipe and applications}},  {\em JHEP} {\bf 0209} (2002) 042
  [\href{http://arXiv.org/abs/hep-th/0205051}{{\tt hep-th/0205051}}].

\bibitem{Berti:2009kk}
E.~Berti, V.~Cardoso and A.~O. Starinets, {\it {Quasinormal modes of black
  holes and black branes}},  {\em Class.Quant.Grav.} {\bf 26} (2009) 163001
  [\href{http://arXiv.org/abs/0905.2975}{{\tt 0905.2975}}].

\bibitem{Balasubramanian:2012tu}
V.~Balasubramanian, A.~Bernamonti, B.~Craps, V.~Ker{\"a}nen, E.~Keski-Vakkuri
  {\em et.~al.}, {\it {Thermalization of the spectral function in strongly
  coupled two dimensional conformal field theories}},  {\em JHEP} {\bf 1304}
  (2013) 069 [\href{http://arXiv.org/abs/1212.6066}{{\tt 1212.6066}}].

\bibitem{Ishii:2015gia}
T.~Ishii, E.~Kiritsis and C.~Rosen, {\it {Thermalization in a Holographic
  Confining Gauge Theory}},  \href{http://arXiv.org/abs/1503.07766}{{\tt
  1503.07766}}.

\bibitem{David:2015xqa}
J.~R. David and S.~Khetrapal, {\it {Thermalization of Green functions and
  quasinormal modes}},  \href{http://arXiv.org/abs/1504.04439}{{\tt
  1504.04439}}.

\bibitem{Bhattacharya:2013bna}
J.~Bhattacharya and T.~Takayanagi, {\it {Entropic Counterpart of Perturbative
  Einstein Equation}},  {\em JHEP} {\bf 1310} (2013) 219
  [\href{http://arXiv.org/abs/1308.3792}{{\tt 1308.3792}}].

\bibitem{Fischler:2012ca}
W.~Fischler and S.~Kundu, {\it {Strongly Coupled Gauge Theories: High and Low
  Temperature Behavior of Non-local Observables}},  {\em JHEP} {\bf 1305}
  (2013) 098 [\href{http://arXiv.org/abs/1212.2643}{{\tt 1212.2643}}].

\bibitem{Grumiller:2008va}
D.~Grumiller and P.~Romatschke, {\it {On the collision of two shock waves in
  AdS5}},  {\em JHEP} {\bf 08} (2008) 027
  [\href{http://arXiv.org/abs/0803.3226}{{\tt 0803.3226}}].

\bibitem{Gubser:2008pc}
S.~S. Gubser, S.~S. Pufu and A.~Yarom, {\it {Entropy production in collisions
  of gravitational shock waves and of heavy ions}},  {\em Phys.Rev.} {\bf D78}
  (2008) 066014 [\href{http://arXiv.org/abs/0805.1551}{{\tt 0805.1551}}].

\bibitem{Chesler:2010bi}
P.~M. Chesler and L.~G. Yaffe, {\it {Holography and colliding gravitational
  shock waves in asymptotically AdS$_5$ spacetime}},  {\em Phys.Rev.Lett.} {\bf
  106} (2011) 021601 [\href{http://arXiv.org/abs/1011.3562}{{\tt 1011.3562}}].

\bibitem{Wu:2011yd}
B.~Wu and P.~Romatschke, {\it {Shock wave collisions in AdS5: approximate
  numerical solutions}},  {\em Int.J.Mod.Phys.} {\bf C22} (2011) 1317--1342
  [\href{http://arXiv.org/abs/1108.3715}{{\tt 1108.3715}}].

\bibitem{Casalderrey-Solana:2013aba}
J.~Casalderrey-Solana, M.~P. Heller, D.~Mateos and W.~van~der Schee, {\it {From
  full stopping to transparency in a holographic model of heavy ion
  collisions}},  {\em Phys.Rev.Lett.} {\bf 111} (2013) 181601
  [\href{http://arXiv.org/abs/1305.4919}{{\tt 1305.4919}}].

\bibitem{boyd2001chebyshev}
J.~Boyd, {\em Chebyshev and Fourier Spectral Methods: Second Revised Edition}.
\newblock Dover Books on Mathematics. Dover Publications, 2001.

\bibitem{Trefethen:2000:SMM:343368}
L.~N. Trefethen, {\em Spectral Methods in MatLab}.
\newblock Society for Industrial and Applied Mathematics, Philadelphia, PA,
  USA, 2000.

\end{thebibliography}

\providecommand{\href}[2]{#2}\begingroup\raggedright\endgroup

\end{document}